\documentclass[journal,10pt,a4paper,onecolumn,draftcls]{IEEEtran}

\usepackage{bm}
\usepackage{dsfont}
\usepackage{mathrsfs}
\usepackage{authblk}
\usepackage{amsmath}
\usepackage{graphicx}
\usepackage{amssymb}
\usepackage{epstopdf}
\usepackage{subfigure}
\usepackage{textcomp}
\usepackage{color}
\usepackage{upgreek}
\usepackage{amsmath}
\usepackage{hyperref}

\newtheorem{lemma}{Lemma}

\begin{document}
\title{Non-Orthogonal Unicast and Broadcast Transmission via Joint Beamforming and LDM in Cellular Networks\thanks{Part of this work was presented at the IEEE Global Communications Conference (Globecom), Washington, D.C., Dec. 2016. \cite{7842028}}}

\author[$\dag$]{Junlin Zhao}
\author[$\dag$]{Deniz G\"und\"uz}
\author[*]{Osvaldo Simeone}
\author[$\ddag$]{David G\'omez-Barquero}
\affil[$\dag$]{Department of Electrical and Electronic Engineering, Imperial College London, London, UK}
\affil[*]{Department of Informatics, King's College London, London, UK}
\affil[$\ddag$]{Institute of Telecommunications and Multimedia Applications, Universitat Polit\`ecnica de Val\`encia, Val\`encia, Spain
\authorcr Email: j.zhao15@imperial.ac.uk, d.gunduz@imperial.ac.uk, osvaldo.simeone@kcl.ac.uk, dagobar@iteam.upv.es}


\maketitle

\begin{abstract}

Limited bandwidth resources and higher energy efficiency requirements motivate incorporating multicast and broadcast transmission into the next-generation cellular network architectures, particularly for multimedia streaming applications.
Layered division multiplexing (LDM), a form of non-orthogonal multiple access (NOMA), can potentially improve unicast throughput and broadcast coverage with respect to traditional orthogonal frequency division multiplexing (FDM) or time division multiplexing (TDM), by simultaneously using the same frequency and time resources for multiple unicast or broadcast transmissions. In this paper, the performance of LDM-based unicast and broadcast transmission in a cellular network is studied by assuming a single frequency network (SFN) operation for the broadcast layer, while allowing arbitrarily clustered cooperation among the base stations (BSs) for the transmission of unicast data streams. Beamforming and power allocation between unicast and broadcast layers, the so-called \emph{injection level} in the LDM literature, are optimized with the aim of minimizing the sum-power under constraints on the user-specific unicast rates and on the common broadcast rate. The effects of imperfect channel coding and imperfect channel state information (CSI) are also studied to gain insights into robust implementation in practical systems. The non-convex optimization problem is tackled by means of successive convex approximation (SCA) techniques. Performance upper bounds are also presented by means of the $\rm{S}$-procedure followed by semidefinite relaxation (SDR). Finally, a dual decomposition-based solution is proposed to facilitate an efficient distributed implementation of LDM where the optimal unicast beamforming vectors can be obtained locally by the cooperating BSs. Numerical results are presented, which show the tightness of the proposed bounds and hence the near-optimality of the proposed solutions.
\end{abstract}

\IEEEpeerreviewmaketitle

\section{Introduction}
With the growing demand for multimedia streaming applications, research efforts to incorporate multicast and broadcast transmission into the cellular network architecture have intensified in recent years. In 3G networks, multimedia broadcast multicast services (MBMS) was introduced to support new point-to-multipoint radio bearers and multicast capability in the core network \cite{4114796}. However, due to its reduced capacity, which did not meet the requirement of mass media services, MBMS has never been deployed commercially. The broadcast extension of 4G LTE is named evolved MBMS (eMBMS), commercially known as LTE Broadcast \cite{Qualcomm}.

Following many field trials worldwide, the first commercial deployment of eMBMS was launched in South Korea in 2014. eMBMS provides full integration and seamless transition between broadcast and unicast modes \cite{6179779}, and significant performance improvement with respect to MBMS, thanks to the higher and more flexible data rates provided by the LTE architecture. Furthermore, it also allows single frequency network (SFN) operation across different cells as in digital television broadcasting, since the LTE waveform is OFDM-based. While it is commonly accepted that eMBMS, in its current form, needs further enhancements to be adopted as a successful commercial platform for TV broadcasting \cite{6823652}, it has been proposed as a converged platform in the UHF band for TV and mobile broadband \cite{6701362}, \cite{CellTV}. For eMBMS TV services, a study has been carried out within 3GPP in 2015 for application scenarios and use cases, as well as for potential requirements and improvements \cite{3GPP}. In 2017, advances have been published in the 3GPP Release 14, including standardization of radio interfaces between mobile network operators and broadcasters and the possibility for free-to-air reception, which is an essential feature for broadcasting TV programs over mobile networks \cite{3gpp2}.
While the standardization and evolvement of point-to-multipoint transmission are primarily led by multimedia broadcasting services, point-to-multipoint transmission techniques have also been adopted in LTE-Advanced Pro for emerging use cases including vehicular to everything (V2X), Internet of things (IoT) and machine-type communication (MCC) \cite{8334920}.

LTE Broadcast entails a reduction in system capacity for unicast services, since eMBMS and unicast services are multiplexed in time in different sub-frames. Superposition coding, a form of non-orthogonal multiple access (NOMA), was proposed in \cite{4557052} to improve unicast throughput and broadcast coverage with respect to traditional orthogonal frequency division multiplexing (FDM) or time division multiplexing (TDM), by simultaneously using the same frequency and time resources for multiple unicast or broadcast transmissions. Superposition coding has been adopted in the next-generation TV broadcasting US standard ATSC 3.0 \cite{7383281} under the name layer division multiplexing (LDM) \cite{7329951}.

At the cost of an increased complexity at the receivers, which need to perform interference cancellation by decoding the generic broadcast content prior to decoding the unicast content, LDM may provide significant gains especially when the superposed signals exhibit large disparities in terms of signal-to-noise-plus-interference ratio (SINR). This is expected to be the case for multiplexing broadcast and unicast services. In fact, the unicast throughput is limited by intercell interference; and hence, increasing the transmit unicast power across the network does not necessarily improve the unicast SINR. In contrast, broadcast does not suffer from intercell interference in an SFN, and increasing the broadcast power results in an increased SINR. This not only helps improve the reliability of the broadcast layer, but it also reduces the interference on the unicast messages as the broadcast layer can be decoded and cancelled more reliably. A performance comparison of LDM with TDM/FDM for unequal error protection in broadcast systems in the absence of multicell interference from an information theoretic perspective can be found in \cite{compare}.

In this paper, we study the performance of non-orthogonal unicast and broadcast transmission in a cellular network via LDM, in order to demonstrate and quantify its benefits compared to orthogonal transmission methods, i.e., TDM and FDM. We assume an SFN operation for the broadcast layer, while allowing arbitrarily clustered cooperation for the unicast data streams. Cooperative transmission for broadcast traffic, and potentially also for unicast data streams, takes place by means of distributed beamforming at multi-antenna base stations. To better account for potential practical impairments, and to evaluate the robustness of LDM in real systems, we also consider imperfections in channel state information (CSI) through an additive error model. Beamforming and power allocation between unicast and broadcast layers, and the so-called injection level in the LDM literature (see, e.g., \cite{compare}), are optimized with the aim of minimizing the sum-power under constraints on the user-specific unicast rates and the common broadcast rate. The optimization of orthogonal transmission via TDM/FDM is also studied for comparison, and the corresponding nonconvex optimization problems are tackled by means of successive convex approximation (SCA) techniques \cite{7776948}, as well as through the calculation of performance upper bounds by means of the S-procedure followed by semidefinite relaxation (SDR) \cite{Boyd:2004:CO:993483}.
Finally, we also present an efficient distributed implementation of the proposed LDM system based on dual decomposition. The dual decomposition based-algorithm allows each cluster of cooperating BSs to optimize their beamforming vectors locally with limited information exchange.

Finally, we also present an efficient distributed implementation of the proposed LDM system based on the dual decomposition method. The dual decomposition based-algorithm allows each cluster of BSs cooperating to transmit a unicast message to obtain their beamforming vector locally with limited information exchange. A completely distributed implementation is not viable due to the presence of the broadcast layer, whose beamforming vector needs to be determined centrally at one of the BSs or in the cloud; however, local computation of the unicast beamforming vectors allows exploiting the computation resources distributed across the network, which can help parallelize these computations.

With regards to previous work, the optimization of the beamforming vectors in multicell systems has been investigated in \cite{Xiang} and \cite{8064707}, where the base station in each cell multicasts one or more data streams to the specified given groups of in-cell users. The coexistence of broadcast and unicast traffic is studied in \cite{7374738}, where the surplus of degrees-of-freedom provided by massive MIMO systems is leveraged to broadcast data to a group of users whose CSI is not available, without creating interference to conventional unicast users.
Recently, the rate splitting technique is considered in \cite{RateSpliting} to construct the unicast and multicast messages, which are then transmitted through joint beamforming.
Robust coordinated beamforming in a multicell network with imperfect CSI is studied in \cite{5590310}, where the optimization problem is solved by a second-order cone program after relaxing the worst-case SINR requirement. The same problem is also studied in \cite{Zheng2008}, \cite{CIT-069}, and \cite{6156468}, where the infinitely many constraints introduced due to the imperfect channel estimation are tackled by the $\rm{S}$-procedure.

Distributed implementations of multigroup multicast beamforming have also been a focus in the literature. A dual decomposition-based scheme has been proposed in \cite{Tolli_dual} by creating consensus over inter-cell interference terms between all the BSs. In \cite{Tervo_primal_ADMM}, a primal decomposition-based algorithm and an alternating direction method of multipliers (ADMM)-based algorithm have been proposed for the SDR version of the original problem. In \cite{7874154}, instead of directly dealing with the relaxed problem, the authors proposed to apply ADMM for each of the convexified SCA problems, obtaining a doule-loop scheme.
In \cite{6156468}, an ADMM-based algorithm is proposed for a distributed solution of the problem with imperfect CSI after relaxing the original problem with $\rm{S}$-procedure.

The rest of this paper is organized as follows. Section II introduces the system model and the problem formulation.
In Section III, the characterized problem is tackled by using the $\rm{S}$-procedure and the SCA technique. Dual decomposition-based distributed algorithms for both TDM and LDM are developed in Section IV. Numerical results are presented in Section V, followed by the conclusions in Section VI.

\section{System Model and Problem Formulation}
In this section, we present the model of the joint unicast and broadcast transmission system under study, by highlighting orthogonal and non-orthogonal multiplexing schemes. For both schemes, we formulate a power minimization problem under user quality of service (QoS) constraints.

\subsection{System Model}

We investigate downlink transmission in a cellular network that serves both unicast and broadcast traffic. Specifically, we focus on a scenario in which a dedicated unicast data stream is to be delivered to each user, while there is a common broadcast data stream intended for all the users. A more general broadcast traffic model, in which distinct data streams are sent to different subsets of users, could be included in the analysis at the cost of a more cumbersome notation, but will not be further pursued in this paper.

\begin{figure}[!pt]
\centering
\includegraphics[width=3in]{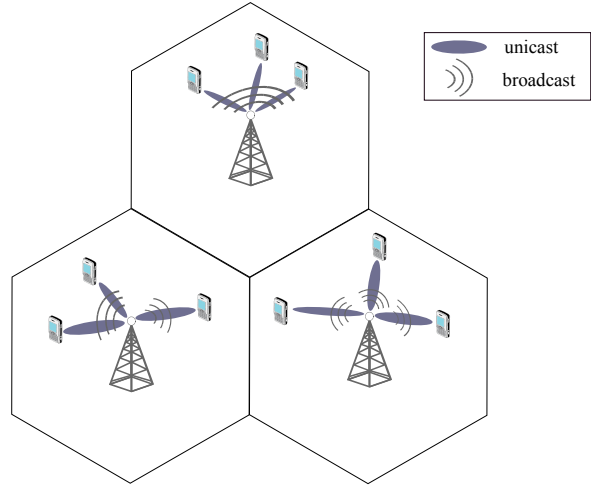}
\caption{Illustration of a multicell network with \(N\)=3 cells and \(K=3\) users in each cell with simultaneous unicast and broadcast transmission.}
\label{illustration}
\end{figure}

As illustrated in Fig.~\ref{illustration}, the network is comprised of \(N\) cells, each consisting of a base station (BS) with \(M\) antennas and \(K\) single-antenna mobile users. The notation (\(n,k\)) identifies the \(k\)-th user in cell \(n\). All BSs cooperate via joint beamforming for the broadcast stream to all the users, while an arbitrary cluster \(\mathcal{C}_{n,k}\) of BSs cooperate for the unicast transmission to user \((n,k)\). Accordingly, all the BSs have access to the broadcast data stream, while only the BSs in cluster \(\mathcal{C}_{n,k}\) are informed about the unicast data stream to be delivered to user \((n,k)\). Note that, non-cooperative unicast transmission, whereby each BS serves only the users in its own cell, can be obtained as a special case when \(\mathcal{C}_{n,k}=\{n\}\), for all users (\(n,k\)). Similarly, fully cooperative unicast transmission is obtained when \(\mathcal{C}_{n,k}=\{1,\ldots,N\}\), for all users (\(n,k\)). We denote the set of users whose unicast messages are available at BS \(i\) as
\begin{eqnarray}
\mathcal{U}_i = \{ (n,k) \ | \ i \in \mathcal{C}_{n,k} \}.
\end{eqnarray}

We assume frequency-flat quasi-static complex channels, and define \(\bm{h}_{i,n,k} \in \mathds{C}^{M \times 1}\) as the channel vector from the BS in cell \(i\) to user (\(n,k\)). We use the notation \(s_{n,k}^U\) to denote an encoded unicast symbol intended for user \((n,k)\), and \(s^B\) to represent an encoded broadcast symbol.
The signal received by user \((n,k)\) at any given channel use can then be written as
\begin{eqnarray}\label{eq:y}
y_{n,k} = \sum\limits_{i=1}^N \bm{h}_{i,n,k}^H \bm{x}_i + n_{n,k},
\end{eqnarray}
where \(\bm{x}_i \in \mathds{C}^{M \times 1}\) is the symbol transmitted by BS \(i\), and \(n_{n,k} \sim \mathcal{CN}(0,\sigma_{n,k}^2)\) is the additive white Gaussian noise. We assume that both the intended and the interference signals at each user are in perfect synchronization without inter-symbol interference.

In practice, BSs have to operate with imperfect CSI. In Frequency Division Duplex (FDD) systems, it may arise from errors in downlink training-based CSI estimation, limited resolution in CSI feedback links, or from delays in CSI acquisition over fading channels, while in Time Division Duplex (TDD) systems, CSI errors are caused by impairments in channel estimation or imperfect channel reciprocity (see \cite{CIT-069} and references therein). As common in the literature, we model the CSI uncertainty with an additive error by setting
\begin{align}\label{eq:channel}
\bm{h}_{i,n,k} = \hat{\bm{h}}_{i,n,k} + \bm{e}_{i,n,k},
\end{align}
where \(\hat{\bm{h}}_{i,n,k} \in \mathds{C}^{M \times 1}\) is the estimated complex channel vector from cell \(i\) to user \((n,k)\) available at the BSs, and \(\bm{e}_{i,n,k} \in \mathds{C}^{M \times 1}\) is the additive channel error. We consider a bounded error, which is typically used to model CSI imperfections resulting from quantization error due to feedback links of limited capacity. Hence, the set of channel vectors from BS $i$ to user $(n,k)$ can be defined as
\begin{align}\label{eq:general_uncertaitny}
\mathcal{H}_{i,n,k} = \{  \bm{h}_{i,n,k}: \bm{h}_{i,n,k}=\hat{\bm{h}}_{i,n,k} + \bm{e}_{i,n,k},~\bm{e}_{i,n,k}^H \bm{Q}_{i,n,k} \bm{e}_{i,n,k} \leq 1  \},~~ \forall i,n,k,
\end{align}
where $\bm{Q}_{i,n,k}$ is a known positive definite matrix. Accordingly, the structure of the uncertainty set of the quantization error vectors is known at the transmitters.

In what follows, we will consider two modes of transmission, namely orthogonal transmission via TDM and non-orthogonal transmission via LDM, where the former will serve as a benchmark to evaluate the potential performance gains from the LDM scheme.

\subsubsection{TDM}

We first consider the standard TDM approach based on the orthogonal transmission of unicast and broadcast signals. Note that orthogonalization can also be realized by means of other multiplexing schemes such as FDM, yielding the same mathematical formulation. With TDM, each transmission slot of duration \(T\) channel uses is divided into two subslots: a subslot of duration \(T_0\) channel uses for unicast transmission, and a subslot of duration \(T-T_0\) for broadcast transmission. Therefore, the signal \(\bm{x}_i\) transmitted by cell \(i\) can be written as
\begin{eqnarray}\label{eq:tdm_x}
\bm{x}_i=
\begin{cases}
\sum\limits_{(n,k) \in \mathcal{U}_i} \bm{w}_{i,n,k}^U s_{n,k}^U & \text{for} \ \ 0 \leq t < T_0 \\
\bm{w}_i^B s^B & \text{for} \ \ T_0 \leq t < T
\end{cases},
\end{eqnarray}
where \(\bm{w}_{i,n,k}^U \in \mathds{C}^{M \times 1}\) represents the unicast beamforming vector applied at the BS in cell \(i\) towards user \((n,k)\), and \(\bm{w}_i^B \in \mathds{C}^{M \times 1}\) is the broadcast beamforming vector applied at the same BS.

The received signal \(y_{n,k}\) at user (\(n,k\)) can be expressed as
\begin{align}\label{eq:tdm_y}
y_{n,k}=\begin{cases}
\Big( \sum\limits_{i \in \mathcal{C}_{n,k}} \bm{h}_{i,n,k}^H \bm{w}_{i,n,k}^U \Big) s_{n,k}^U + z_{n,k} + n_{n,k}& \text{for} \ 0 \leq t < T_0\\
\Big(\sum\limits_{i=1}^{N} \bm{h}_{i,n,k}^H \bm{w}_i^B\Big) s^B + n_{n,k} & \text{for} \ T_0 \leq t < T
\end{cases},
\end{align}
where
\begin{eqnarray}\label{eq:i}
z_{n,k}=\sum\limits_{(p,q)\neq(n,k)} \bigg(\sum_{i \in \mathcal{C}_{p,q}} \bm{h}_{i,n,k}^H \bm{w}_{i,p,q}^U\bigg) s_{p,q}^U
\end{eqnarray}
denotes the interference at user (\(n,k\)).

\subsubsection{LDM}

In LDM, the transmitted signal \(\bm{x}_i\) from the BS in cell \(i\) is the superposition of the broadcast and unicast signals for the entire time slot, which can be written as
\begin{eqnarray}\label{eq:ldm_x}
\bm{x}_i=\bm{w}_i^B s^B + \sum\limits_{(n,k) \in \mathcal{U}_i} \bm{w}_{i,n,k}^U s_{n,k}^U \quad \text{for} \ 0 \leq t \leq T,
\end{eqnarray}
for all channel uses in an entire time slot, i.e., for \(0 \leq t \leq T\). We note that the power ratio between broadcast and unicast, which is referred to as the $\emph{injection~level}$ (IL) in the literature (see, e.g., \cite{compare}), can be obtained as
\begin{eqnarray}
\text{IL} = 10~\text{log}_{10}\frac{P^B}{P^U},
\end{eqnarray}
where \(P^B=\sum_{i=1}^N ||\bm{w}_i^B||^2\) is the total broadcast power, and \(P^U=\sum_{i=1}^N\sum_{(n,k) \in \mathcal{U}_i} ||\bm{w}_{i,n,k}^U||^2\) is the total unicast power.
The received signal at user (\(n,k\)) is given by
\begin{eqnarray}\label{eq:ldm_y}
y_{n,k} & = &\Big( \sum\limits_{i=1}^N \bm{h}_{i,n,k}^H \bm{w}_i^B \Big) s^B + \Big( \sum\limits_{i \in \mathcal{C}_{n,k}} \bm{h}_{i,n,k}^H \bm{w}_{i,n,k}^U \Big) s_{n,k}^U + z_{n,k} + n_{n,k}, \quad \text{for} \ 0 \leq t \leq T,
\end{eqnarray}
where \(z_{n,k}\) is the interference as defined in (\ref{eq:i}).

\subsection{Problem Formulation}

The power minimization problem for the above systems can be expressed in the following form:
\begin{subequations}\label{eq:optimization}
\begin{flalign}
\mathop{\text{min}}_{\{\bm{w}_i^B\},\{\bm{w}_{i,n,k}^U\}} \ \ & \sum\limits_{i=1}^N \ \Big(||\bm{w}_i^B||^2 + \sum\limits_{(n,k) \in \mathcal{U}_i} ||\bm{w}_{i,n,k}^U||^2\Big)\\
\text{s.t.} \ \ & \mathop{\text{min}}_{\mathcal{H}}~\text{SINR}_{n,k}^B \ge \gamma^B, \ \forall n,k, \label{eq:BC}\\
&\mathop{\text{min}}_{\mathcal{H}}~\text{SINR}_{n,k}^U \ge \gamma_{n,k}^U, \ \forall n,k,\label{eq:UC}
\end{flalign}
\end{subequations}
where the explicit expressions for the SINRs at user (\(n,k\)) for broadcast and unicast transmissions, namely \(\text{SINR}_{n,k}^B\) and \(\text{SINR}_{n,k}^U\) will be given below for TDM and LDM separately.
The constraints in (\ref{eq:BC}) and (\ref{eq:UC}) are imposed on the worst-case SINRs for all possible channel realizations in the set $\mathcal{H}=\prod_{i,n,k} \mathcal{H}_{i,n,k}$.
Note that, since all the users receive the same broadcast signal, we have enforced a common broadcast QoS requirement. In contrast, the unicast SINR requirements are allowed to be user-dependent.

\subsubsection{TDM}
From the expression of the received signal in (\ref{eq:tdm_y}), we derive the SINR for the broadcast layer in TDM for user \((n,k)\) as
\begin{eqnarray}\label{eq:TDM SINR BC}
\text{SINR}_{n,k}^{B\text{-TDM}} = \frac{ |\bm{h}_{n,k}^H \bm{w}^B|^2 } {\sigma_{n,k}^2},
\end{eqnarray}
where \(\bm{h}_{n,k} = [\bm{h}_{1,n,k}^T,\ldots,\bm{h}_{N,n,k}^T]^T~\in~\mathds{C}^{NM \times 1}\) is the aggregated channel vector from all the BSs to user \((n,k)\). All broadcast beamforming vectors are similarly aggregated into the vector \(\bm{w}^B = [\bm{w}_1^{B^T},\ldots,\bm{w}_N^{B^T}]^T \in \mathds{C}^{NM \times 1}\). The SINR for the unicast layer is instead given as
\begin{eqnarray}\label{eq:TDM SINR UC}
\text{SINR}_{n,k}^{U\text{-TDM}} = \frac{ |\bm{h}_{n,k}^{{(n,k)}^H} \bm{w}_{n,k}^U|^2 } { \sum\limits_{(p,q)\neq(n,k)} |\bm{h}_{n,k}^{{(p,q)}^H} \bm{w}_{p,q}^U|^2 + \sigma_{n,k}^2 },
\end{eqnarray}
where \(\bm{h}_{n,k}^{(p,q)} = [\bm{h}_{i,n,k}^{T}]^T_{i \in \mathcal{C}_{p,q}}\) is the aggregated channel vector to user \((n,k)\) from all the BSs in cluster \(\mathcal{C}_{p,q}\) of BSs that serve user \((p,q)\), and \(\bm{w}_{n,k}^U = [\bm{w}_{i,n,k}^{U^T}]^T_{i \in \mathcal{C}_{n,k}}\) is similarly defined as the aggregate unicast beamforming vector for user \((n,k)\) from all the BSs in cluster \(\mathcal{C}_{n,k}\).

We observe that the SINR targets \(\gamma_{n,k}^{U\text{-TDM}}\) and \(\gamma^{B\text{-TDM}}\) for unicast and broadcast traffic can be obtained from the corresponding transmission rates  \(R_{n,k}^U\) and \(R^B\), respectively, as
\begin{eqnarray}\label{eq:TDM UC target}
\frac{T_0}T \log_2(1+\gamma_{n,k}^{U\text{-TDM}}) = R_{n,k}^U,
\end{eqnarray}
and
\begin{eqnarray}\label{eq:TDM BC target}
\frac{T-T_0}T \log_2(1+\gamma^{B\text{-TDM}}) = R^B.
\end{eqnarray}

\subsubsection{LDM}
With LDM, the broadcast layer, which is intended for all the users and usually has a higher SINR, is decoded first by treating unicast signals as noise, as in \cite{4557052}. The users decode their unicast data streams after canceling the decoded broadcast message. The broadcast SINR in LDM for user \((n,k)\) is hence obtained from the received signal (\ref{eq:ldm_y}) as follows
\begin{eqnarray}\label{eq:LDM SINR BC}
\text{SINR}_{n,k}^{B\text{-LDM}}=
\frac{|\bm{h}_{n,k}^H \bm{w}^B|^2}{\sum\limits_{(p,q)} |\bm{h}_{n,k}^{{(p,q)}^H} \bm{w}_{p,q}^U|^2 + \sigma_{n,k}^2},
\end{eqnarray}
while the unicast SINR is the same as TDM given in (\ref{eq:TDM SINR UC}), i.e.,
\begin{eqnarray}\label{eq:LDM SINR UC}
\text{SINR}_{n,k}^{U\text{-LDM}} = \text{SINR}_{n,k}^{U\text{-TDM}}.
\end{eqnarray}
Similarly to TDM, SINR thresholds for unicast and broadcast can be obtained from the transmission rates \(R_{n,k}^U\) and \(R^B\), respectively, as
\begin{eqnarray}\label{eq:LDM UC target}
\log_2(1+\gamma_{n,k}^{U\text{-LDM}}) = R_{n,k}^U,
\end{eqnarray}
and
\begin{eqnarray}\label{eq:LDM BC target}
\log_2(1+\gamma^{B\text{-LDM}}) = R^B.
\end{eqnarray}

In \cite{7842028}, a performance lower bound on the power minimization problem is obtained by standard semidefinite relaxation (SDR), assuming that perfect CSI is available at all the BSs. In this paper, the problem formulation incorporates CSI uncertainty in (\ref{eq:BC}) and (\ref{eq:UC}) by imposing constraints on the worst-case performance over all possible channel realizations on the optimization problem. The formulated worst-case quadratically-constrained quadratic program (QCQP) is intractable due to the induced additional constraints on the CSI error vectors.
Nevertheless, the uncertainty due to CSI errors can be tackled by applying the \(\rm{S}\)-procedure as in \cite{CIT-069}, as a result of which SDR can be employed as in the perfect CSI case to obtain a lower bound on the optimal solution.
Furthermore, an achievable beamformer design under the worst-case SINR constraints will be obtained based on SCA, and its performance will be compared with the obtained lower bound.

\section{Bounds on the minimum total power}
The optimization problem formulated in (\ref{eq:optimization}) is nonconvex due to the QoS constraints in (\ref{eq:BC}) and (\ref{eq:UC}). Therefore, there are in general no numerical solution techniques with guaranteed convergence to a global optimal solution. In this section, we will present numerical tools to obtain lower and upper bounds on the minimum total transmit power.

\subsection{Lower Bound via \(\rm{S}\)-Procedure}

The optimization problem in (\ref{eq:optimization}) contains an infinite number of constraints in (\ref{eq:BC}) and (\ref{eq:UC}), thus it is intractable. To address this issue, \(\rm{S}\)-procedure \cite{Boyd:2004:CO:993483} will be adopted to derive an equivalent but tractable problem formulation.
Following the CSI error model in (\ref{eq:general_uncertaitny}) we can form the aggregated CSI error vector $\bm{e}_{n,k}$ for user \((n,k)\) consistent with the aggregated channel vector $\bm{h}_{n,k}$, and define the relaxed set of possible channel vectors to user $(n,k)$ as:
\begin{align}\label{eq:combined_error}
\mathcal{H}_{n,k} \triangleq \{  \bm{h}_{n,k}: \bm{h}_{n,k}=\hat{\bm{h}}_{n,k} + \bm{e}_{n,k},~\bm{e}_{n,k}^H \bm{Q}_{n,k} \bm{e}_{n,k} \leq 1  \},
\end{align}
where
\begin{align}
\bm{Q}_{n,k}\triangleq\frac{1}{N}
\begin{bmatrix}
\bm{Q}_{1,n,k} & \ & \bm{0}\\
\ & \ddots & \ \\
\bm{0} & \ & \bm{Q}_{N,n,k}
\end{bmatrix}.
\end{align}
It is noted that the set of possible channel vectors in (\ref{eq:combined_error}) is a relaxed version of the original set given in (\ref{eq:general_uncertaitny}). For reference, we present the \(\rm{S}\)-procedure in the following lemma for completeness.

\begin{lemma}[S-procedure]\label{s-lemma}
Let $f_i(\bm{x}) \triangleq \bm{x}^H\bm{F}_i\bm{x} + \bm{g}_i^H\bm{x} + \bm{x}^H\bm{g}_i+c_i,~\text{for}~i=0,1$, where $\bm{F}_i\in\mathds{C}^{NM \times NM}$ is Hermitian semidefinite, $\bm{g}\in \mathds{C}^{{NM} \times 1}$, and $c_i \in \mathds{R}$, then $f_1(\bm{x}) \leq 0$ for all $\bm{x}$ satisfying $f_0(\bm{x}) \leq 0$ holds if and only if there exists a $\lambda \geq 0$ such that
\begin{align}
\left[
\begin{array}{cc}
\bm{F}_1 & \bm{g}_1\\
\bm{g}_1^H & c_1
\end{array}
\right]
\preceq
\lambda \left[
\begin{array}{cc}
\bm{F}_0 & \bm{g}_0\\
\bm{g}_0^H & c_0
\end{array}
\right].
\end{align}
\end{lemma}

\subsubsection{TDM}
The constraint for the broadcast layer in (\ref{eq:BC}) can be rewritten as
\begin{eqnarray}\label{eq:BC_reformulation}
(\hat{\bm{h}}_{n,k}^H+\bm{e}_{n,k}^H) \bm{W}^B (\hat{\bm{h}}_{n,k}+\bm{e}_{n,k}) \geq \sigma_{n,k}^2\gamma_{n,k}^B,~\text{for}~\forall \bm{e}_{n,k}^H \bm{Q}_{n,k} \bm{e}_{n,k} \leq 1,
\end{eqnarray}
where $\bm{W}^B\triangleq\bm{w}^B\bm{w}^{B^H}$.
By applying the \(\rm{S}\)-procedure, the worst-case SINR constraint in (\ref{eq:BC}) can be recast as
\begin{align}\label{eq:S_TDM_BC}
\left[
\begin{array}{cc}
\bm{W}^B & \bm{W}^B \hat{\bm{h}}_{n,k}\\
\hat{\bm{h}}_{n,k}^H \bm{W}^B & \frac{1}{\gamma_{n,k}^B}\hat{\bm{h}}_{n,k}^H\bm{W}^B\hat{\bm{h}}_{n,k} - \sigma_{n,k}^2
\end{array}
\right]
+
\lambda_{n,k}^B
\left[
\begin{array}{cc}
\bm{Q}_{n,k} & \bm{0}\\
\bm{0}^T & -1
\end{array}
\right]{}
\succeq 0,
\end{align}
for some \(\lambda_{n,k}^B \ge 0, \ \forall n,k\).
Accordingly to Lemma~\ref{s-lemma}, the constraints on the unicast transmissions in (\ref{eq:UC}) can be written as
\begin{align}\label{eq:UC_reformulation_1}
(\hat{\bm{h}}_{n,k}+\bm{e}_{n,k})^H \Big(\frac{1}{\gamma_{n,k}^U}\bm{T}_{n,k}^T \bm{W}_{n,k}^U \bm{T}_{n,k} - \sum\limits_{(p,q)\neq(n,k)} \bm{T}_{p,q}^T \bm{W}_{p,q}^U \bm{T}_{p,q}\Big) (\hat{\bm{h}}_{n,k}+\bm{e}_{n,k}) \geq \sigma_{n,k}^2,~\text{for}~\forall \bm{e}_{n,k}^H \bm{Q}_{n,k} \bm{e}_{n,k} \leq 1,
\end{align}
where $\bm{W}_{n,k}^U\triangleq\bm{w}_{n,k}^U\bm{w}_{n,k}^{U^H}$, and $\bm{T}_{p,q}$ is a constructed block matrix of dimension $|\mathcal{C}_{p,q}|\times N$ such that $\bm{h}_{n,k}^{(p,q)}=\bm{T}_{p,q} \bm{h}_{n,k}$.
Following the $\rm{S}$-procedure, the worst-case SINR constraint for the unicast layer can be recast as
\begin{align}\label{eq:S_TDM_UC}
\left[
\begin{array}{lr}
\bm{V}_{n,k} & \bm{V}_{n,k} \hat{\bm{h}}_{n,k}\\
\hat{\bm{h}}_{n,k}^H \bm{V}_{n,k} & \hat{\bm{h}}_{n,k}^H\bm{V}_{n,k}\hat{\bm{h}}_{n,k} - \sigma_{n,k}^2
\end{array}
\right]
+
\lambda_{n,k}^U
\left[
\begin{array}{cc}
\bm{Q}_{n,k} & \bm{0}\\
\bm{0}^T & -1
\end{array}
\right]
\succeq 0, \ \forall n,k,
\end{align}
for some \(\lambda_{n,k}^U \ge 0, \ \forall n,k\), where \(\bm{V}_{n,k}\) is defined as
\begin{align}
\bm{V}_{n,k} \triangleq \frac{1}{\gamma_{n,k}^U}\bm{T}_{n,k}^T \bm{W}_{n,k}^U \bm{T}_{n,k} - \sum\limits_{(p,q)\neq(n,k)} \bm{T}_{p,q}^T \bm{W}_{p,q}^U \bm{T}_{p,q}.
\end{align}
Following these transforms and definitions, the problem in (\ref{eq:optimization}) can be relaxed to a tractable semidefinite program by dropping the rank constraints on matrices \(\bm{W}^B\) and \(\bm{W}_{n,k}^U\).
Specifically, for TDM, the relaxed problem after SDR is given by
\begin{subequations}\label{eq:S_TDM}
\begin{flalign}
\mathop{\text{min}}_{\bm{W}^B,\{\bm{W}_{n,k}^U\},\{\lambda_{n,k}^B\},\{\lambda_{n,k}^U\}} \ \ & \text{tr}(\bm{W}^B) + \sum\limits_{n=1}^N\sum\limits_{k=1}^K \text{tr}(\bm{W}_{n,k}^U)\\
\text{s.t.} \ \ & (\ref{eq:S_TDM_BC}) \text{~and~}  (\ref{eq:S_TDM_UC}),\\
&\lambda_{n,k}^B \ge 0, \lambda_{n,k}^U \ge 0, \ \forall n,k.
\end{flalign}
\end{subequations}

\subsubsection{LDM}

Similar to the analysis in TDM, the constraint on the broadcast transmission in LDM can be equivalently written as
\begin{align}\label{eq:S_LDM_BC}
\left[
\begin{array}{lr}
\bm{U} & \bm{U} \hat{\bm{h}}_{n,k}\\
\hat{\bm{h}}_{n,k}^H \bm{U} & \hat{\bm{h}}_{n,k}^H\bm{U}\hat{\bm{h}}_{n,k} - \sigma_{n,k}^2
\end{array}
\right]
+
\lambda_{n,k}^B
\left[
\begin{array}{cc}
\bm{Q}_{n,k} & \bm{0}\\
\bm{0}^T & -1
\end{array}
\right]
\succeq 0,
\end{align}
where \(\lambda_{n,k}^U \ge 0, \ \forall n,k\), and \(\bm{U}\) is defined as
\begin{align}
\bm{U} \triangleq \frac{1}{\gamma_{n,k}^B}\bm{W}^B - \sum\limits_{(p,q)} \bm{T}_{p,q}^T \bm{W}_{p,q}^U \bm{T}_{p,q}.
\end{align}
The unicast constraint in LDM can be reformulated as in (\ref{eq:S_TDM_UC}), hence the relaxed problem after dropping the rank-1 constraints on matrices \(\bm{W}^B\) and \(\bm{W}_{n,k}^U\) is obtained as follows:
\begin{subequations}\label{eq:S_LDM}
\begin{flalign}
\mathop{\text{min}}_{\bm{W}^B,\{\bm{W}_{n,k}^U\},\{\lambda_{n,k}^B\},\{\lambda_{n,k}^U\}} \ \ & \text{tr}(\bm{W}^B) + \sum\limits_{n=1}^N\sum\limits_{k=1}^K \text{tr}(\bm{W}_{n,k}^U)\\
\text{s.t.} \ \ & (\ref{eq:S_TDM_UC}) \text{~and~} (\ref{eq:S_LDM_BC}),\\
&\lambda_{n,k}^B \ge 0, \lambda_{n,k}^U \ge 0, \ \forall n,k.
\end{flalign}
\end{subequations}
As the rank-1 constraint has been dropped in (\ref{eq:S_TDM}) and (\ref{eq:S_LDM}), the corresponding optimal solutions provide lower bounds on the optimal solutions of the original problems in (\ref{eq:optimization}). Note that, under perfect CSI, i.e., $\bm{e}_{i,n,k}=\bm{0}$, the problem formulation in (11) boils down to the one presented in \cite{7842028}, and the solution obtained by first applying the $\rm{S}$-procedure is equal to that obtained directly by SDR.

\subsection{Upper Bound via SCA}

Instead of adopting Gaussian randomization \cite{5447068} to obtain a feasible (achievable) beamforming scheme, we leverage the SCA method \cite{7776948} to obtain an achievable beamformer, which yields an upper bound on the minimum required power. In particular, by rewriting the nonconvex QoS constraints as the difference of convex (DC) functions, the SCA algorithm reduces to the conventional convex-concave procedure \cite{6788812}. We remark that the SCA scheme is known to converge to a stationary point of the original problem \cite{7776948}.

In order to apply the SCA approach, each nonconvex constraint in (\ref{eq:optimization}) will be expressed as
\begin{eqnarray}\label{eq:SCA_1}
g(\bm{w}) = g^+(\bm{w}) - g^-(\bm{w}) \le 0,
\end{eqnarray}
where \(g^+(\bm{w})\) and \(g^-(\bm{w})\) are both convex functions on the set of all beamforming vectors \(\bm{w}\). Then a convex upper bound is obtained by linearizing the nonconvex part around any given vector \(\bm{u}\), yielding the stricter constraint on the solution \(\bm{w}\) as
\begin{eqnarray}\label{eq:SCA_2}
\tilde{g}(\bm{w};\bm{u}) \triangleq g^+(\bm{w}) - g^-(\bm{u}) - \bigtriangledown_{\bm{w}} g^-(\bm{u})^T(\bm{w}-\bm{u}) \le 0.
\end{eqnarray}

\subsubsection{TDM}
The constraint in (11b) on the broadcast layer can be approximated and replaced by the following tighter constraint:
\begin{align}\label{eq:TDM BC reformulation 1}
|\hat{\bm{h}}_{n,k}^H \bm{w}^B| - |\bm{e}_{n,k}^H\bm{w}^B| \geq \sqrt{\gamma^B}\sigma_{n,k}~\text{for}~\forall~\bm{e}_{n,k}^H\bm{Q}_{n,k}\bm{e}_{n,k} \leq 1,
\end{align}
which can be further tightened as:
\begin{align}\label{eq:TDM_SCA_BC}
|\hat{\bm{h}}_{n,k}^H \bm{w}^B| - \Vert\bm{Q}_{n,k}^{-\frac{1}{2}} \bm{w}^B\Vert \geq \sqrt{\gamma^B}\sigma_{n,k},
\end{align}
since $|\bm{e}_{n,k}^H\bm{w}^B| \leq \Vert\bm{Q}_{n,k}^{-\frac{1}{2}} \bm{w}^B\Vert$ holds for the CSI error vectors $\bm{e}_{n,k}$ as we have $\bm{e}_{n,k} \in \{\bm{Q}_{n,k}^{-\frac{1}{2}}\bm{u}~|~\Vert\bm{u}\Vert \leq 1\}$.

The constraint in (\ref{eq:TDM_SCA_BC}) is in the DC form, for which SCA can be adopted to obtain an iterative algorithm which converges to a stationary point of the original problem.
The constraint at the $\nu$-th iteration of the SCA algorithm is given by
\begin{align}\label{eq:TDM BC reformulation den relax SCA}
\sqrt{\gamma^B}\sigma_{n,k} + \Vert\bm{Q}_{n,k}^{-\frac{1}{2}} \bm{w}^B\Vert + |\hat{\bm{h}}_{n,k}^H \bm{w}^B(\nu)| -2 \frac{\hat{\bm{h}}_{n,k}^H \hat{\bm{h}}_{n,k}\bm{w}^{B^H}(\nu)}{\vert\hat{\bm{h}}_{n,k}^H\bm{w}^{B}(\nu)\vert}\bm{w}^B\leq 0,~\forall n,k.
\end{align}
Also, the constraint in (\ref{eq:UC}) for the unicast transmission can be tightened by considering the worst-case SINR, i.e.,
\begin{eqnarray}
\frac{ \mathop{\text{min}}\limits_{\mathcal{H}}~|\bm{h}_{n,k}^{{(n,k)}^H} \bm{w}_{n,k}^U|^2 } { \mathop{\text{max}}\limits_{\mathcal{H}} \sum\limits_{(p,q)\neq(n,k)} |\bm{h}_{n,k}^{{(p,q)}^H} \bm{w}_{p,q}^U|^2 + \sigma_{n,k}^2 } \geq \gamma^U_{n,k},~\text{for}~\forall n,k,
\end{eqnarray}
which can then be replaced equivalently by the following set of constraints:
\begin{subequations}\label{eq:TDM UC reformulation}
\begin{align}
&\mathop{\text{max}}\limits_{\mathcal{H}} \ |\bm{h}_{n,k}^{{(p,q)}^H} \bm{w}_{p,q}^U| \leq \beta_{n,k}^{(p,q)},~\forall n,k,\forall(p,q)\neq(n,k),\label{eq:TDM UC reformulation numerator}\\
&\mathop{\text{min}}\limits_{\mathcal{H}} \ |\bm{h}_{n,k}^{{(n,k)}^H} \bm{w}_{n,k}^U| \geq t^U_{n,k},\label{eq:TDM UC reformulation den}\\
&\gamma^U_{n,k}\Big(\sum\limits_{(p,q)\neq(n,k)}\beta_{n,k}^{{(p,q)}^2}+\sigma_{n,k}^2\Big) - t_{n,k}^{U^2} \leq 0,\label{eq:TDM UC reformulation quot}
\end{align}
\end{subequations}
where $\{t_{n,k}^U\}$ and $\{\beta_{n,k}^{(p,q)}\}$ are auxiliary variables. Note that $\beta_{n,k}^{(p,q)}$ indicates the interference power from BSs in the cluster $\mathcal{C}_{p,q}$ to user $(n,k)$, and $t_{n,k}^U$ indicates the received unicast power at user $(n,k)$.
The constraint in (\ref{eq:TDM UC reformulation numerator}) and (\ref{eq:TDM UC reformulation den}) can be further relaxed by
\begin{align}\label{eq:TDM UC reformulation numerator relax}
|\hat{\bm{h}}_{n,k}^{{(p,q)}^H} \bm{w}_{p,q}^U| + |\bm{Q}_{n,k}^{{(p,q)}^{-1/2}}\bm{w}_{p,q}^U| \leq \beta_{n,k}^{(p,q)},~\forall n,k,\forall(p,q)\neq(n,k),
\end{align}
and
\begin{align}\label{eq:TDM UC reformulation den relax}
t_{n,k}^U + \Vert \bm{Q}_{n,k}^{{(n,k)}^{-1/2}} \bm{w}^U_{n,k}\Vert - |\hat{\bm{h}}_{n,k}^{{(n,k)^H}} \bm{w}^U_{n,k}| \leq 0,
\end{align}
respectively, where $\bm{Q}_{n,k}^{{(p,q)}^{-1/2}} = \bm{Q}_{n,k}^{{-1/2}}\bm{T}_{p,q}$.
According to (\ref{eq:SCA_1}) and (\ref{eq:SCA_2}), in the SCA algorithm, the corresponding constraints in the $\nu$-th iteration for (\ref{eq:TDM UC reformulation quot}) and (\ref{eq:TDM UC reformulation den relax}) can be written as
\begin{align}\label{eq:TDM UC reformulation quot SCA}
\gamma^U_{n,k}\Big(\sum\limits_{(p,q)\neq(n,k)}\beta_{n,k}^{{(p,q)}^2}+\sigma_{n,k}^2\Big) + t_{n,k}^{U^2}(\nu) - 2t_{n,k}^U(\nu)t_{n,k}^U\leq 0, \ \forall n,k,
\end{align}
and
\begin{align}\label{eq:TDM UC reformulation den relax SCA}
t_{n,k}^U + \Vert\bm{Q}_{n,k}^{(n,k)^{-\frac{1}{2}}}\bm{w}_{n,k}^U\Vert + |\bm{h}_{n,k}^{(n,k)^H}\bm{w}_{n,k}^U(\nu)| -2 \frac{\hat{\bm{h}}_{n,k}^{(n,k)^H}\hat{\bm{h}}_{n,k}^{(n,k)}\bm{w}_{n,k}^{U^H}(\nu)}{\vert\hat{\bm{h}}_{n,k}^{(n,k)^H}\bm{w}_{n,k}^{U}(\nu)\vert}\bm{w}_{n,k}^U \leq 0, \ \forall n,k,
\end{align}
respectively.

Due to the fact that the feasible convexified constraints in (\ref{eq:TDM BC reformulation den relax SCA}), (\ref{eq:TDM UC reformulation numerator relax}), (\ref{eq:TDM UC reformulation quot SCA}) and (\ref{eq:TDM UC reformulation den relax SCA}) are stricter than the original constraints in (\ref{eq:optimization}), the solution obtained at each iteration is feasible for the original problem (\ref{eq:optimization}) as long as a feasible initial point is available. When the stopping criterion is satisfied, we take the last iteration as the solution of the SCA algorithm. Please refer to Table~\ref{tab:SCA Algorithm} for an algorithmic description of the SCA approach.

\begin{table}[!tp]\caption{SCA Algorithm}\label{tab:SCA Algorithm}
\centering
\begin{tabular}{l}\\ \hline
STEP 0: Set \(\nu=1\). Set a step size \(\mu\).\\
\quad\quad\quad\ \ \ Initialize \(\bm{w}^B(1)\) and \(\bm{w}_{n,k}^U(1)\) with feasible values\\
STEP 1: If a stopping criterion is satisfied, then STOP\\
STEP 2: Set \(\bm{w}^B(\nu+1)=\bm{w}^B(\nu) + \mu (\bm{w}^B-\bm{w}^B(\nu))\),\\
\quad\quad\quad\ \ \(\bm{w}_{n,k}^U(\nu+1)=\bm{w}_{n,k}^U(\nu) + \mu (\bm{w}_{n,k}^U-\bm{w}_{n,k}^U(\nu))\),\\
\quad\quad\quad\ \ where \(\{\bm{w}^B\}\) and \(\{\bm{w}_{n,k}^U\}\) are obtained as solutions\\
\quad\quad\quad\ \ of problems (\ref{eq:opt_TDM_SCA}) for TDM and (\ref{eq:opt_LDM_SCA}) for LDM\\
STEP 3: Set \(\nu=\nu+1\), and go to STEP 1\\
\hline
\end{tabular}
\end{table}

When obtaining the numerical results in the next section, initialization of the SCA algorithm is carried out based on the solution \(\{\bm{W}^B\}\) and \(\{\bm{W}_{n,k}^U\}\) obtained from the $\rm{S}$-procedure. Specifically, we perform a rank-1 reduction of matrices \(\{\bm{W}^B\}\) and \(\{\bm{W}_{n,k}^U\}\), obtaining vectors \(\{\bm{w}^B\}\) and \(\{\bm{w}_{n,k}^U\}\), respectively, as the largest principal component. These vectors are then scaled with the smallest common factor \(t\), which is evaluated through line search, to satisfy constraints (\ref{eq:BC}) and (\ref{eq:UC}), yielding the initial points \(\{\bm{w}^B(1)\}\) and \(\{\bm{w}_{n,k}^U(1)\}\) for SCA. If a feasible value for \(t\) is not found through a line search, then the SCA method is considered to be infeasible. Further discussion on this point can be found in Section V.

As a summary, the relaxed version of the problem for (\ref{eq:optimization}) in TDM in the SCA form is given as
\begin{subequations}\label{eq:opt_TDM_SCA}
\begin{flalign}
\mathop{\text{min}}_{\bm{w}^B,\{\bm{w}_{n,k}^U\},\{\beta_{n,k}^{(p,q)}\},\{t_{n,k}^U\}} \ \ & \Vert\bm{w}^B\Vert^2 + \sum\limits_{(n,k)} \Vert\bm{w}_{n,k}^U\Vert^2\\
\text{s.t.} \ \ & ~(\ref{eq:TDM BC reformulation den relax SCA}),~(\ref{eq:TDM UC reformulation numerator relax}),~(\ref{eq:TDM UC reformulation quot SCA}),~\text{and}~(\ref{eq:TDM UC reformulation den relax SCA}).
\end{flalign}
\end{subequations}

\subsubsection{LDM}
Similarly to the TDM approach, the constraint in (\ref{eq:BC}) can be relaxed as the worst-case SINR constraint, i.e.,
\begin{eqnarray}
\frac{\mathop{\text{min}}\limits_{\mathcal{H}} \ |\bm{h}_{n,k}^H \bm{w}^B|^2}{\mathop{\text{max}} \limits_{\mathcal{H}}\sum\limits_{(p,q)} |\bm{h}_{n,k}^{{(p,q)}^H} \bm{w}_{p,q}^U|^2 + \sigma_{n,k}^2} \geq \gamma^B,
\end{eqnarray}
which is then replaced by the following equivalent constraints:
\begin{subequations}\label{eq:LDM BC reformulation}
\begin{align}
&\mathop{\text{max}}\limits_{\mathcal{H}}~|\bm{h}_{n,k}^{{(p,q)}^H} \bm{w}_{p,q}^U| \leq \beta_{n,k}^{(p,q)},\label{eq:LDM BC reformulation numerator}\\
&\mathop{\text{min}}\limits_{\mathcal{H}} \ |\bm{h}_{n,k}^H \bm{w}^B| \geq t_{n,k}^B,\label{eq:LDM BC reformulation den}\\
&\gamma^U_{n,k}\big(\sum\limits_{(p,q)}\beta_{n,k}^{{(p,q)}^2}+\sigma_{n,k}^2\Big) - t_{n,k}^{B^2} \leq 0\label{eq:LDM BC reformulation quot}
\end{align}
\end{subequations}
for all $n,k$, where \(\{t_{n,k}^B\}\) are auxiliary variables indicating the received broadcast power at user $(n,k)$.
Similarly to the relaxation we adopt for the TDM case, the constraint in (\ref{eq:LDM BC reformulation numerator}) can be relaxed as
\begin{align}\label{eq:LDM BC reformulation numerator relax}
|\hat{\bm{h}}_{n,k}^{{(p,q)}^H} \bm{w}_{p,q}^U| + |\bm{Q}_{n,k}^{{(p,q)}^{-1/2}}\bm{w}_{p,q}^U| \leq \beta_{n,k}^{(p,q)},\ \forall n,k,p,q
\end{align}
for all $n,k$.
The constraint in (\ref{eq:LDM BC reformulation den}) can be relaxed as
\begin{align}\label{eq:LDM BC reformulation den relax}
t_{n,k}^B + \Vert\bm{Q}_{n,k}^{-\frac{1}{2}} \bm{w}^B\Vert - |\hat{\bm{h}}_{n,k}^H \bm{w}^B| \leq 0,
\end{align}
which is in the convex-concave form.
According to (\ref{eq:SCA_1}) and (\ref{eq:SCA_2}), in the SCA algorithm, the corresponding constraints in the $\nu$-th iteration for (\ref{eq:LDM BC reformulation quot}) and (\ref{eq:LDM BC reformulation den relax}) can be written as
\begin{align}\label{eq:LDM BC reformulation quot SCA}
\gamma^B_{n,k}\Big(\sum\limits_{(p,q)}\beta_{n,k}^{{(p,q)}^2}+\sigma_{n,k}^2\Big) + t_{n,k}^{B^2}(\nu) - 2t_{n,k}^B(\nu)t_{n,k}^B\leq 0, \ \forall n,k,
\end{align}
and
\begin{align}\label{eq:LDM UC reformulation den relax SCA}
t_{n,k}^B + \Vert\bm{Q}_{n,k}^{{-\frac{1}{2}}}\bm{w}^B\Vert + |\bm{h}_{n,k}^{H}\bm{w}^B(\nu)| -2 \frac{\hat{\bm{h}}_{n,k}^{H}\hat{\bm{h}}_{n,k}\bm{w}^{B^H}(\nu)}{\vert\hat{\bm{h}}_{n,k}^{H}\bm{w}^{B}(\nu)\vert}\bm{w}^B \leq 0, \ \forall n,k,
\end{align}
respectively.
As a summary, the relaxed version of the (\ref{eq:optimization}) for LDM in the SCA form is given as
\begin{subequations}\label{eq:opt_LDM_SCA}
\begin{flalign}
\mathop{\text{min}}_{\bm{w}^B,\{\bm{w}_{n,k}^U\},\{\beta_{n,k}^{(p,q)}\},\{t_{n,k}^B\},\{t_{n,k}^U\}} \ \ & \Vert\bm{w}^B\Vert^2 + \sum\limits_{(n,k)} \Vert\bm{w}_{n,k}^U\Vert^2\\
\text{s.t.} \ \ & (\ref{eq:TDM UC reformulation quot SCA}),~(\ref{eq:TDM UC reformulation den relax SCA}),~(\ref{eq:LDM BC reformulation numerator relax}), (\ref{eq:LDM BC reformulation quot SCA}),~\text{and}~(\ref{eq:LDM UC reformulation den relax SCA}).
\end{flalign}
\end{subequations}

\section{Dual Decomposition-based Distributed Optimization}\label{s:ADMM}

In this section, we propose a distributed algorithm to solve the SCA problem in (\ref{eq:opt_LDM_SCA}) using dual decomposition as in \cite{7776948}.
In particular, while the broadcast beamforming vector $\bm{w}^B$ is designed at a central node that gathers full CSI between all the BSs and the users, the optimization of unicast beamforming vectors $\{\bm{w}_{n,k}^U\}$ is offloaded to the processing unit of the corresponding cluster $\mathcal{C}_{n,k}$, which can be located at one of the BSs within the cluster. This distributed implementation is made possible by the fact that the optimization of $\{\bm{w}_{n,k}^U\}$ can be decomposed into $NK$ independent subproblems, and the processing unit of each cluster $\mathcal{C}_{n,k}$ can calculate $\bm{w}^U_{n,k}$ locally, but still optimally, based only on local CSI, in addition to certain limited information exchange with other clusters.

The benefits of this distributed implementation are as follows. First of all, it reduces the computational requirements on the unique central processing unit, which will in turn reduce the overall latency due to computation delay. Also, transmitting all the CSI back to a unique central unit may lead to increased CSI uncertainty as the CSI may need to be compressed at higher rates to be communicated to a single node.

This means that the CSI error at the central processing unit may be higher as compared with the local BSs; and, therefore, the local computation of the beamforming vectors may be more efficient. In our formulation here, for simplicity, we consider the same CSI error variance for both the broadcast and unicast beamforming optimization problems.
Finally, in the absence of a broadcast message destined for the whole network, all computations can be carried out locally at the cluster heads.

For clarity, to start, we reproduce the optimization in (\ref{eq:opt_LDM_SCA}) as
\begin{subequations}\label{eq:opt_LDM_SCA_reproduce}
\begin{flalign}
\mathop{\text{min}}_{\bm{w}^B,\{\bm{w}_{n,k}^U\},\{\beta_{n,k}^{(p,q)}\},\{t_{n,k}^B\},\{t_{n,k}^U\}} \ & \Vert\bm{w}^B\Vert^2 + \sum\limits_{(n,k)} \Vert\bm{w}_{n,k}^U\Vert^2\\
\text{s.t.} \ \ &|\hat{\bm{h}}_{n,k}^{{(p,q)}^H} \bm{w}_{p,q}^U| + \Vert\bm{Q}_{n,k}^{{(p,q)}^{-1/2}}\bm{w}_{p,q}^U\Vert \leq \beta_{n,k}^{(p,q)}, \ \forall n,k,p,q\label{eq:opt_LDM_SCA_1}\\
&\gamma^U_{n,k} \left(\sum\limits_{(p,q)\neq(n,k)}\beta_{n,k}^{{(p,q)}^2}+\sigma_{n,k}^2 \right) + t_{n,k}^{U^2}(\nu) - 2t^U_{n,k}(\nu)t_{n,k}^U\leq 0, \ \forall n,k,\label{eq:opt_LDM_SCA_2}\\
&t^U_{n,k} + \Vert\bm{Q}_{n,k}^{(n,k)^{-\frac{1}{2}}}\bm{w}_{n,k}^U\Vert + |\hat{\bm{h}}_{n,k}^{(n,k)^H}\bm{w}_{n,k}^U(\nu)| -2 \frac{\hat{\bm{h}}_{n,k}^{(n,k)^H}\hat{\bm{h}}_{n,k}^{(n,k)}\bm{w}_{n,k}^{U^H}(\nu)}{\vert\hat{\bm{h}}_{n,k}^{(n,k)^H}\bm{w}_{n,k}^{U}(\nu)\vert}\bm{w}_{n,k}^U \leq 0, \ \forall n,k,\label{eq:opt_LDM_SCA_3}\\
&\gamma^B_{n,k} \left(\sum\limits_{(p,q)}\beta_{n,k}^{{(p,q)}^2} + \sigma_{n,k}^2 \right) + t_{n,k}^{B^2}(\nu) - 2t^B_{n,k}(\nu)t_{n,k}^B\leq 0, \ \forall n,k,\label{eq:opt_LDM_SCA_4}\\
&t^B_{n,k} + \Vert\bm{Q}_{n,k}^{{-\frac{1}{2}}}\bm{w}^B\Vert + |\hat{\bm{h}}_{n,k}^{H}\bm{w}^B(\nu)| -2 \frac{\hat{\bm{h}}_{n,k}^{H}\hat{\bm{h}}_{n,k}\bm{w}^{B^H}(\nu)}{\vert\hat{\bm{h}}_{n,k}^{H}\bm{w}^{B}(\nu)\vert}\bm{w}^B \leq 0, \ \forall n,k.\label{eq:opt_LDM_SCA_5}
\end{flalign}
\end{subequations}
We now introduce Lagrangian multipliers $\bm{\lambda} \triangleq\{\lambda_{n,k}^{(p,q)}\},\bm{\mu}\triangleq\{\mu_{n,k}\},\bm{\kappa}\triangleq\{\kappa_{n,k}\},\bm{\xi}\triangleq\{\xi_{n,k}\},\bm{\rho}\triangleq\{\rho_{n,k}\}$ for the constraints in (\ref{eq:opt_LDM_SCA_1})-(\ref{eq:opt_LDM_SCA_5}), respectively, and define $\bm{z}\triangleq\left(  \bm{w}^B,\{\bm{w}^U_{n,k}\},\{\beta_{n,k}^{(p,q)}\},\{t_{n,k}^B\},\{t_{n,k}^U\}\right)$. Then the Lagrangian of (\ref{eq:opt_LDM_SCA_reproduce}) can then be obtained as
\begin{align}
\mathcal{L}\left( \bm{\lambda},\bm{\mu}, \bm{\kappa},\bm{\xi},\bm{\rho},\bm{z};\bm{z}(\nu) \right) &=  \mathcal{L} _{\bm{w}^B}\left(  \bm{\rho},\bm{w}^B;\bm{w}^B(\nu) \right) + \sum\limits_{n,k}  \mathcal{L} _{\bm{w}_{n,k}^U}\left(\bm{\lambda}_{n,k},\kappa_{n,k},\bm{w}_{n,k}^U;\bm{w}_{n,k}^U(\nu)\right)
+  \sum\limits_{n,k}  \mathcal{L} _{\bm{\beta}_{n,k}}\left(  \bm{\lambda}_{n,k},\mu_{n,k},\xi_{n,k},\bm{\beta}_{n,k} \right)\nonumber\\
&+  \sum\limits_{n,k}   \mathcal{L} _{t_{n,k}^U}\left( \mu_{n,k},\kappa_{n,k},t_{n,k}^U;t_{n,k}^U(\nu) \right)
+  \sum\limits_{n,k}  \mathcal{L} _{t_{n,k}^B} \left( \xi_{n,k},\rho_{n,k},t_{n,k}^B;t_{n,k}^B(\nu) \right),
\end{align}
where
\begin{subequations}
\begin{align}
&\mathcal{L}_{\bm{w}^B} \left(  \bm{\rho},\bm{w}^B;\bm{w}^B(\nu) \right)\triangleq \Vert\bm{w}^B\Vert^2 + \sum\limits_{n,k}\rho_{n,k}  \Vert\bm{Q}_{n,k}^{{-\frac{1}{2}}}\bm{w}^B\Vert -2\sum\limits_{n,k}\rho_{n,k} \frac{\hat{\bm{h}}_{n,k}^{H}\hat{\bm{h}}_{n,k}\bm{w}^{B^H}(\nu)}{\vert\hat{\bm{h}}_{n,k}^{H}\bm{w}^{B}(\nu)\vert}\bm{w}^B,
\end{align}
\begin{align}
\mathcal{L}_{\bm{w}_{n,k}^U}\left(\bm{\lambda}^{(n,k)},\kappa_{n,k},\bm{w}_{n,k}^U;\bm{w}_{n,k}^U(\nu)\right)\triangleq& \Vert\bm{w}_{n,k}^U\Vert^2 + \sum\limits_{p,q}\lambda_{p,q}^{(n,k)}\left( |\hat{\bm{h}}_{p,q}^{{(n,k)}^H} \bm{w}_{n,k}^U| + |\bm{Q}_{p,q}^{{(n,k)}^{-1/2}}\bm{w}_{n,k}^U|\right)\nonumber\\+&\kappa_{n,k} \Vert\bm{Q}_{n,k}^{(n,k)^{-\frac{1}{2}}}\bm{w}_{n,k}^U\Vert -2\kappa_{n,k} \frac{\hat{\bm{h}}_{n,k}^{(n,k)^H}\hat{\bm{h}}_{n,k}^{(n,k)}\bm{w}_{n,k}^{U^H}(\nu)}{\vert\hat{\bm{h}}_{n,k}^{(n,k)^H}\bm{w}_{n,k}^{U}(\nu)\vert}\bm{w}_{n,k}^U,
\end{align}
\begin{align}
\mathcal{L}_{\bm{\beta}_{n,k}}\left(  \bm{\lambda}_{n,k},\mu_{n,k},\xi_{n,k},\bm{\beta}_{n,k} \right)\triangleq- \sum\limits_{p,q}\lambda_{n,k}^{(p,q)}\beta_{n,k}^{(p,q)} + \mu_{n,k}\gamma_{n,k}^U \sum\limits_{(p,q)\neq(n,k)}\beta_{n,k}^{{(p,q)}^2} + \xi_{n,k}\gamma_{n,k}^B \sum\limits_{(p,q)}\beta_{n,k}^{{(p,q)}^2},
\end{align}
\begin{align}
\mathcal{L}_{t_{n,k}^U}\left( \mu_{n,k},\kappa_{n,k},t_{n,k}^U;t_{n,k}^U(\nu) \right) \triangleq -2\mu_{n,k}  t_{n,k}^U(\nu)t^U_{n,k} + \kappa_{n,k} t_{n,k}^U ,\\
\mathcal{L}_{t_{n,k}^B}\left( \xi_{n,k},\rho_{n,k},t_{n,k}^B;t_{n,k}^B(\nu) \right) \triangleq -2\xi_{n,k}  t_{n,k}^B(\nu)t_{n,k}^B + \rho_{n,k} t_{n,k}^B.
\end{align}
\end{subequations}
The optimization problem in (\ref{eq:opt_LDM_SCA_reproduce}) is strongly convex and satisfies Slater's condition, thus strong duality holds. Therefore, the optimal solution can be obtained by solving its dual problem, which is given by
\begin{subequations}
\begin{align}
\mathop{\text{max}}_{\bm{\lambda},\bm{\mu}, \bm{\kappa},\bm{\xi},\bm{\rho}} \ \ &D \left( \bm{\lambda},\bm{\mu}, \bm{\kappa},\bm{\xi},\bm{\rho} ;\bm{z}(\nu) \right)\\
\text{s.t.} \ \
&~\bm{\lambda}\geq\bm{0},\bm{\mu}\geq\bm{0}, \bm{\kappa}\geq\bm{0},\bm{\xi}\geq\bm{0},\bm{\rho}\geq\bm{0},
\end{align}
\end{subequations}
where the dual function $D\left( \bm{\lambda},\bm{\mu}, \bm{\kappa},\bm{\xi},\bm{\rho} ;\bm{z}(\nu) \right)$ is obtained by minimizing the Lagrangian over the primal variables as
\begin{subequations}\label{eq:dual function}
\begin{align}
    D\left( \bm{\lambda},\bm{\mu}, \bm{\kappa},\bm{\xi},\bm{\rho} ;\bm{z}(\nu) \right) &= \mathop{\text{min}}_{\bm{w}^B}\mathcal{L}_{\bm{w}^B}\left(  \bm{\rho},\bm{w}^B;\bm{w}^B(\nu) \right)\label{eq:dual function_1}\\
    +&
    \sum_{n,k}\mathop{\text{min}}_{\bm{w}^U_{n,k}}\mathcal{L}_{\bm{w}_{n,k}^U}\left(\bm{\lambda}_{n,k},\kappa_{n,k},\bm{w}_{n,k}^U;\bm{w}_{n,k}^U(\nu)\right)\label{eq:dual function_2}\\
    +&\sum_{n,k}\mathop{\text{min}}_{\beta_{n,k}^{(p,q)}}\mathcal{L}_{\{\beta_{n,k}^{(p,q)}\}}\left(  \bm{\lambda}_{n,k},\mu_{n,k},\xi_{n,k},\beta_{n,k}^{(p,q)} \right)\label{eq:dual function_3}\\
    +&\sum_{n,k}\mathop{\text{min}}_{t^U_{n,k}}\mathcal{L}_{t_{n,k}^U}\left( \mu_{n,k},\kappa_{n,k},t_{n,k}^U;t_{n,k}^U(\nu) \right)\label{eq:dual function_4}\\
    +&\sum_{n,k}\mathop{\text{min}}_{t^B_{n,k}}\mathcal{L}_{t_{n,k}^B}\left( \xi_{n,k},\rho_{n,k},t_{n,k}^B;t_{n,k}^B(\nu) \right),\label{eq:dual function_5}
\end{align}
\end{subequations}
yielding the optimal solutions $\hat{\bm{z}}\left(\bm{\lambda},\bm{\mu}, \bm{\kappa},\bm{\xi},\bm{\rho} \right)=\left(\hat{\bm{w}}^B_{n,k},\{\hat{\bm{w}}^U_{n,k}\},\{\hat{\beta}_{n,k}^{(p,q)}\},\{\hat{t}_{n,k}^U\},\{\hat{t}^B_{n,k}\}\right)$.
The optimization over $\bm{w}^U_{n,k},\beta_{n,k}^{(p,q)},t^U_{n,k},t^B_{n,k}$ in (\ref{eq:dual function}) can be decomposed into $NK$ separable subproblems. The dual function $D\left( \bm{\lambda},\bm{\mu}, \bm{\kappa},\bm{\xi},\bm{\rho} ;\bm{z}(\nu) \right)$ is differentiable with its gradient given by
\begin{subequations}\label{eq:gradient_dual}
\begin{align}
&\nabla_{\lambda_{p,q}^{n,k}} D\left( \bm{\lambda},\bm{\mu}, \bm{\kappa},\bm{\xi},\bm{\rho} ;\hat{\bm{z}}(\nu) \right)=|\hat{\bm{h}}_{p,q}^{{(n,k)}^H} \hat{\bm{w}}_{n,k}^U| + \Vert\bm{Q}_{p,q}^{{(n,k)}^{-1/2}}\hat{\bm{w}}_{n,k}^U\Vert - \hat{\beta}_{p,q}^{(n,k)},~\forall p,q,\label{eq:gradient_dual_1}\\
&\nabla_{\mu_{n,k}} D\left( \bm{\lambda},\bm{\mu}, \bm{\kappa},\bm{\xi},\bm{\rho} ;\hat{\bm{z}}(\nu) \right)=\gamma^U_{n,k} \left(\sum\limits_{(p,q)\neq(n,k)}\hat{\beta}_{n,k}^{{(p,q)}^2}+\sigma_{n,k}^2 \right) + \hat{t}_{n,k}^{U^2}(\nu) - 2t^U_{n,k}(\nu)\hat{t}_{n,k}^U\label{eq:gradient_dual_2},\\
&\nabla_{\kappa_{n,k}} D\left( \bm{\lambda},\bm{\mu}, \bm{\kappa},\bm{\xi},\bm{\rho} ;\hat{\bm{z}}(\nu) \right)=\hat{t}^U_{n,k} + \Vert\bm{Q}_{n,k}^{(n,k)^{-\frac{1}{2}}}\hat{\bm{w}}_{n,k}^U\Vert + |\hat{\bm{h}}_{n,k}^{(n,k)^H}\bm{w}_{n,k}^U(\nu)| -2 \frac{\hat{\bm{h}}_{n,k}^{(n,k)^H}\hat{\bm{h}}_{n,k}^{(n,k)}\bm{w}_{n,k}^{U^H}(\nu)}{\vert\hat{\bm{h}}_{n,k}^{(n,k)^H}\bm{w}_{n,k}^{U}(\nu)\vert}\hat{\bm{w}}^U_{n,k}\label{eq:gradient_dual_3},\\
&\nabla_{\xi_{n,k}} D\left( \bm{\lambda},\bm{\mu}, \bm{\kappa},\bm{\xi},\bm{\rho} ;\hat{\bm{z}}(\nu) \right)=\gamma^B_{n,k} \left(\sum\limits_{(p,q)}\hat{\beta}_{n,k}^{{(p,q)}^2} + \sigma_{n,k}^2 \right) + t_{n,k}^{B^2}(\nu) - 2t^B_{n,k}(\nu)\hat{t}_{n,k}^B,\label{eq:gradient_dual_4}\\
&\nabla_{\rho_{n,k}} D\left( \bm{\lambda},\bm{\mu}, \bm{\kappa},\bm{\xi},\bm{\rho} ;\hat{\bm{z}}(\nu) \right)=\hat{t}^B_{n,k} + \Vert\bm{Q}_{n,k}^{{-\frac{1}{2}}}\hat{\bm{w}}^B\Vert + |\hat{\bm{h}}_{n,k}^{H}\hat{\bm{w}}^B(\nu)| -2 \frac{\hat{\bm{h}}_{n,k}^{H}\hat{\bm{h}}_{n,k}\bm{w}^{B^H}(\nu)}{\vert\hat{\bm{h}}_{n,k}^{H}\bm{w}^{B}(\nu)\vert}\hat{\bm{w}}^B,\label{eq:gradient_dual_5}
\end{align}
\end{subequations}
all of which can be computed efficiently in a distributed manner.

Overall, the obtained algorithm is a double-loop scheme. The outer loop consists of the SCA iterations as described in Table~\ref{tab:SCA Algorithm}. In each of the SCA iteration, gradient descent based dual ascent algorithm is adopted. First, the primal variable $\bm{z}^j$ is updated by solving the optimization problems outlined in (\ref{eq:dual function_1})-(\ref{eq:dual function_5}), each of which is solved by solving $NK$ subproblems. Specifically, the update of $\bm{w}_{n,k}^{U^j}$ only requires local CSI, i.e., $\bm{\hat{h}}_{p,q}^{(n,k)}$ for $\forall p,q$, and other local information such as $\bm{\lambda}^{(n,k)}$ and $\kappa_{n,k}$. Similarly, the updates of $t_{n,k}^U$ and $t_{n,k}^B$ only require local information. On the other hand, the update of the networkwide beamforming vector $\bm{w}^{B^j}$ needs full CSI across the network, as well as gathered information $\rho_{n,k}$ from all the clusters. The update of $\bm{\beta}_{n,k}$, which measures the received interference powers at user $(n,k)$ from BSs outside the cluster $\mathcal{C}_{n,k}$, involves the exchange of $\{\lambda_{n,k}^{(p,q)}\}$ from all $p,q$. Once the primal variable is updated, dual variable updates can be executed with the gradient descent method, with gradient given in (\ref{eq:gradient_dual_1})-(\ref{eq:gradient_dual_5}), respectively. Note that the update of dual variables can be performed locally with the message $\bm{w}^{B^j}$ from the central processing unit.
The detailed algorithm description can be found in Table \ref{tab:Dis Algorithm LDM}.

\begin{table}[!tp]\caption{Distributed Algorithm within the $v$-th SCA iteration in LDM}\label{tab:Dis Algorithm LDM}
\centering
\begin{tabular}{l}\\ \hline
STEP 0: Set $j=1$. Initialize dual variables $\bm{\lambda}^0,\bm{\mu}^0, \bm{\kappa}^0,\bm{\xi}^0,\bm{\rho}^0$.\\
STEP 1: If the stopping criterion is satisfied, then STOP\\
STEP 2: At the central node:\\
\quad\quad\quad\quad\quad~~solve (\ref{eq:dual function_1}) to obtain $\bm{w}^{B^{j}}$\\
\quad\quad\quad~~At each cluster $\mathcal{C}_{n,k}$:\\
\quad\quad\quad\quad\quad~~update $\bm{w}_{n,k}^{U^{j}},t^{U^{j}}_{n,k},t_{n,k}^{B^{j}}$ with only local information\\
\quad\quad\quad\quad\quad~~update $\bm{\beta}_{n,k}^{j}$ with $\lambda_{n,k}^{{(p,q)}^{j-1}}$ from $\mathcal{C}_{p,q}$ where $(p,q)\neq(n,k)$\\
STEP3: The central node broadcasts $\bm{w}^{B^j}$ to all the clusters\\
\quad\quad\quad~~Each cluster $\mathcal{C}_{n,k}$ sends $\beta_{n,k}^{{(p,q)}^j}$ to $\mathcal{C}_{p,q}$\\
STEP 4: At each cluster $\mathcal{C}_{n,k}$:\\
\quad\quad\quad\quad\quad~~update $\bm{\lambda}^{{n,k}^j},\mu_{n,k}^j, \kappa_{n,k}^j,\xi_{n,k}^j,\rho_{n,k}^j$ according to (\ref{eq:gradient_dual_1})-(\ref{eq:gradient_dual_5})\\
\quad\quad\quad\quad\quad~~Each cluster $\mathcal{C}_{n,k}$ sends $\lambda_{p,q}^{(n,k)^j}$ to $\mathcal{C}_{p,q}$\\
STEP 4: Set $j=j+1$, and go to STEP 1\\
\hline
\end{tabular}
\end{table}

\section{Simulation Results}

In this section, simulation results are presented to obtain insights into the performance comparison between LDM and TDM for the purpose of transmission of unicast and broadcast services in cellular systems.
Unless stated otherwise, we consider a network comprised of macro-cells, each with $K=3$ single-antenna active users. The radius of each cell is 500 m, and the users are located uniformly around the BS at a distance of 400 m. Each BS is equipped with \(M=3\) antennas. All channel vectors \(\bm{h}_{i,n,k}\) are written as \(\bm{h}_{i,n,k}=\left(10^{-\text{PL}/10}\right)^{1/2}\bm{\tilde{h}}_{i,n,k}\), where the path loss exponent is modeled as \(\text{PL}=148.1+37.6\text{log}_{10}(d_{i,n,k})\), with \(d_{i,n,k}\) denoting the distance (in kilometers) between the \(i\)-th BS and user (\(n,k\)), and \(\bm{\tilde{h}}_{i,n,k}\) denoting an i.i.d. vector accounting for Rayleigh fading of unitary power.
The noise variance is set to \(\sigma_{n,k}^2=-134~\text{dBW}\) for all users (\(n,k\)). Unless stated otherwise, we assume non-cooperative unicast transmission, i.e., each BS is informed only about the unicast data streams of its own users.

\begin{figure}[!tp]
\centering
\includegraphics[width=3.5in]{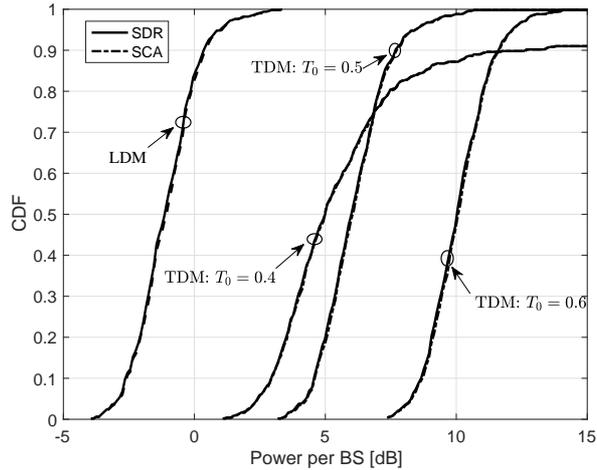}
\caption{The CDF of power consumption per BS with target rates \(R^B\)=3~bps/Hz and \(R^U\)=0.5~bps/Hz.}
\label{CDF}
\end{figure}

\subsection{Perfect CSI}
Initially, we assume perfect CSI at all the BSs in the network. We plot the cumulative distribution function (CDF) of the transmission power per BS for LDM and TDM with \(N=3\) cells in Fig.~\ref{CDF}. For the latter, we consider different values for the fraction of time \(T_0/T\) devoted to unicast traffic. Other values of \(T_0/T\) were seen not to improve the performance. The transmission power per BS is defined as the sum-power divided by the number of BSs. We observe that the curves may represent improper CDFs in the sense that their asymptotic values may be below 1. This gap accounts for the probability of the set of channel realizations in which the problem is found to be infeasible. We refer to the previous section for the assumed definition of infeasibility for SCA, whereas the standard definition is used for the convex problems in (\ref{eq:S_TDM}) and (\ref{eq:S_LDM}) solved using the $\rm{S}$-procedure. Henceforth, we refer to the probability of an infeasible channel realization as the \emph{outage probability}.
\begin{figure}[!htbp]
\centering
\includegraphics[width=3.5in]{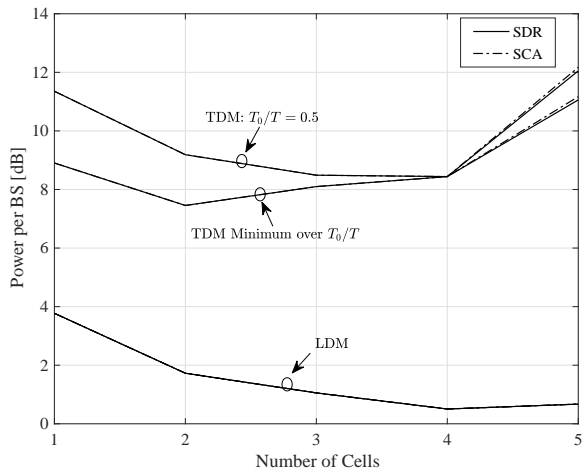}
\caption{Power consumption per BS as a function of the number of cells with target rates \(R^B\)=3 bps/Hz and \(R^U\)=0.5 bps/Hz.}
\label{overall}
\end{figure}

We can observe from Fig.~\ref{CDF} that LDM enables a significant reduction in the transmission power per BS as compared with TDM. In fact, even with an optimized choice of \(T_0/T\), LDM can improve the 95th percentile of the transmitted power per BS by around 7 dB. Another observation is that SCA operates close to the lower bound set by SDR.
Note also that LDM has a significantly lower outage probability than TDM. Finally, we remark that a large value of \(T_0/T\) is beneficial to obtain a lower outage probability in TDM, suggesting that the unicast constraints have more significant impact on the feasibility of the problem due to the need to cope with the mutual interference among unicast data streams. For the rest of this section, the displayed power values correspond to the 95th percentile of the corresponding CDF.

\begin{figure}[!htbp]
\centering
\includegraphics[width=3.5in]{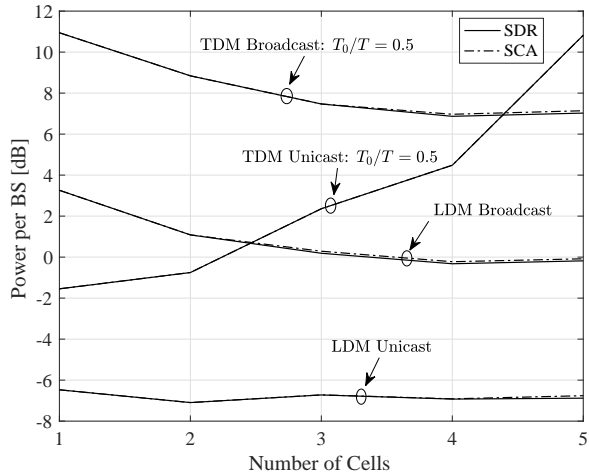}
\caption{Power consumption per BS as a function of the number of cells with target rates \(R^B\)=3 bps/Hz and \(R^U\)=0.5 bps/Hz.}
\label{overallUCBC}
\end{figure}

Next we study the impact of the number of cells on the performance of the system. To this end, Fig.~\ref{overall} and Fig.~\ref{overallUCBC} show the power per BS as a function of the number of cells. Specifically, Fig.~\ref{overall} shows the overall power per BS, while Fig.~\ref{overallUCBC} illustrates separately the power per BS used for the broadcast and unicast layers. Note that in Fig.~\ref{overallUCBC} we fixed \(T_0/T=0.5\), while in Fig.~\ref{overall} we also show the power obtained by selecting, for any number of cells, the value of \(T_0/T\) that minimizes the overall sum-power consumption (obtained by a line search). A key observation from Fig.~\ref{overall} is that the power saving afforded by LDM increases with the number of cells. This gain can be attributed to the following two facts: (\emph{i}) the optimal injection level is high (see Fig.~\ref{overallUCBC}), and hence the broadcast layer requires more power than unicast; and (\emph{ii}) the performance of LDM is enhanced by the presence of more cells broadcasting the same message in the SFN, which increases the broadcast SINR and the broadcast layer can be more easily canceled by the users. The latter fact can be seen from Fig.~\ref{overallUCBC}, in which the required unicast power decreases with the number of cells when using LDM, unlike in TDM. Furthermore, the optimal IL of TDM decreases significantly, also suggesting that TDM is more sensitive to the mutual interference introduced by unicast data streams.

\begin{figure}[!tp]
\centering
\includegraphics[width=3.5in]{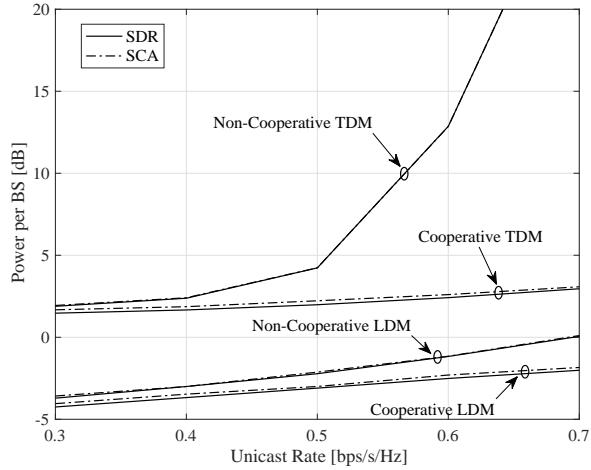}
\caption{Power consumption per BS, separately for the unicast and broadcast signals, for values of unicast rate with \(R^B\)=2~bps/Hz for non-cooperative and fully cooperative schemes.}
\label{Cooperative}
\end{figure}

Fig.~\ref{Cooperative} compares the required power per BS for non-cooperative unicast transmission and for fully cooperative unicast transmission, i.e., clusters \(\mathcal{C}_{n,k}=\{1,\ldots,N\}\) for all users (\(n,k\)). Here we consider a network comprised of \(N=3\) cells, and set \(T_0/T=0.8\) for TDM. From Fig.~\ref{Cooperative}, it can be concluded that a higher unicast rate entails larger power savings by means of cooperative unicast transmission, especially for TDM. It is also worth mentioning that the LDM approach without BS cooperation in unicast transmission can even outperform the fully cooperative TDM approach in certain scenarios, e.g., when the rate for unicast messages is considerably lower than the broadcast rate.

\begin{figure}[!htbp]
\centering
\includegraphics[width=3.5in]{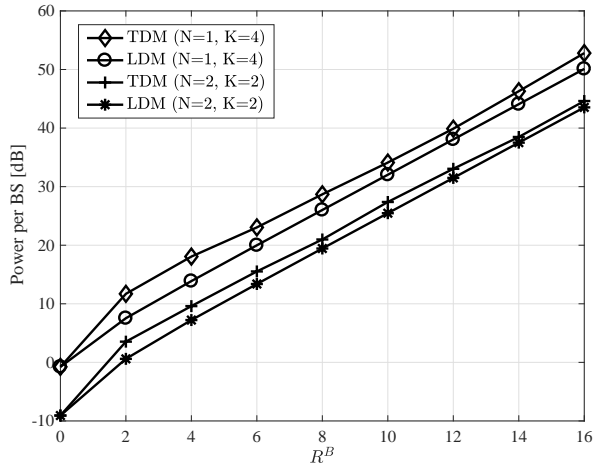}
\caption{Power consumption per BS as a function of the broadcast rate with \(R^U\)=0.5 bps/Hz}
\label{final}
\end{figure}

We present the required power per BS of LDM and TDM as a function of the broadcast rate in Fig.~\ref{final}. The unicast rate is set to $R^U = 0.5$ bps/Hz for all the users. The optimal time allocation $T_0$ in TDM is found by a line search with step size 0.05. When only unicast transmissions exist, i.e., $R^B=0$, both LDM and TDM problems boil down to the multigroup multicast beamforming problem, and have the same performance in terms of power consumption.
When the broadcast message and unicast messages are jointly transmitted, LDM always outperforms TDM in the considered range of broadcast rates. It is also concluded that the performance gain of LDM is larger with a higher user density. 

\begin{figure}[!htbp]
\centering
\includegraphics[width=3.5in]{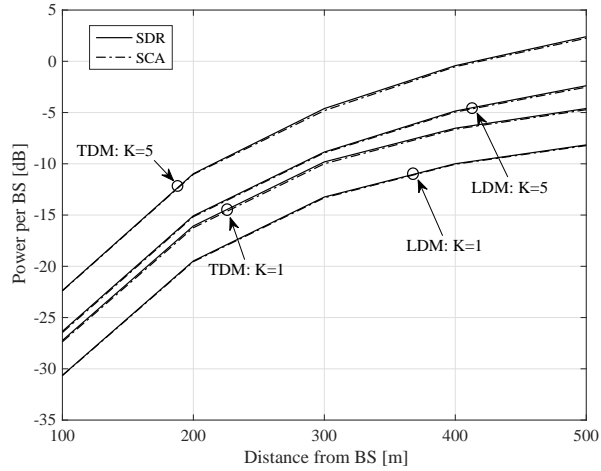}
\caption{Power consumption per BS as a function of the distance between users and BSs with \(R^B\)=1 bps/Hz, \(R^U\)=0.5 bps/Hz, $N=5$, and $M=5$.}
\label{distance}
\end{figure}

Finally, we show the impact of the distance between users and the BS on the performance of TDM and LDM in Fig.~\ref{distance}. Here we consider the network consisting of $N=5$ cells, each with a BS of $M=5$ antennas. The scenarios with $K=1$ and $K=5$ users in each cell are simulated to observe the impact of user density on the performance of the system. It can be seen that LDM always outperforms TDM and has a power gain of around 5 dB in the considered range of distances. It is also observed that LDM can provide the same level of performance for cell-edge users as cell-center users in TDM.

Next, we present the performance comparison between TDM and LDM considering two practical impairments, namely, imperfect channel coding, and imperfect CSI.

\subsection{Imperfect Channel Coding}

To account for the channel coding suboptimality, the SNR gap to capacity for broadcast and unicast layers is introduced as in \cite{compare}. Then, the SINR expressions of the broadcast signal are modified as follows:
\begin{align}
\text{SINR}_{n,k}^{B\text{-TDM}} = \lambda^{B}\frac{ |\bm{h}_{n,k}^H \bm{w}^B|^2 } {\sigma_{n,k}^2}
\end{align}
and
\begin{align}
\text{SINR}_{n,k}^{B\text{-LDM}}=\lambda^B
\frac{|\bm{h}_{n,k}^H \bm{w}^B|^2}{\sum\limits_{(p,q)} |\bm{h}_{n,k}^{{(p,q)}^H} \bm{w}_{p,q}^U|^2 + \sigma_{n,k}^2},
\end{align}
as opposed to (\ref{eq:TDM SINR BC}) and (\ref{eq:LDM SINR BC}) for TDM and LDM, respectively, where $\lambda^B$ is the SNR gap to capacity of the broadcast layer. Similarly, the SINR expressions for the unicast transmission in (\ref{eq:TDM SINR UC}) and (\ref{eq:LDM SINR UC}) are modified to
\begin{align}
\text{SINR}_{n,k}^{U\text{-LDM}} = \text{SINR}_{n,k}^{U\text{-TDM}} = \lambda^U \frac{ |\bm{h}_{n,k}^{{(n,k)}^H} \bm{w}_{n,k}^U|^2 } { \sum\limits_{(p,q)\neq(n,k)} |\bm{h}_{n,k}^{{(p,q)}^H} \bm{w}_{p,q}^U|^2 + \sigma_{n,k}^2 },
\end{align}
where $\lambda^U$ is the SNR gap to capacity for the unicast layer.

\begin{figure}[!tp]
\centering
\subfigure{
\centering
\includegraphics[scale=0.4]{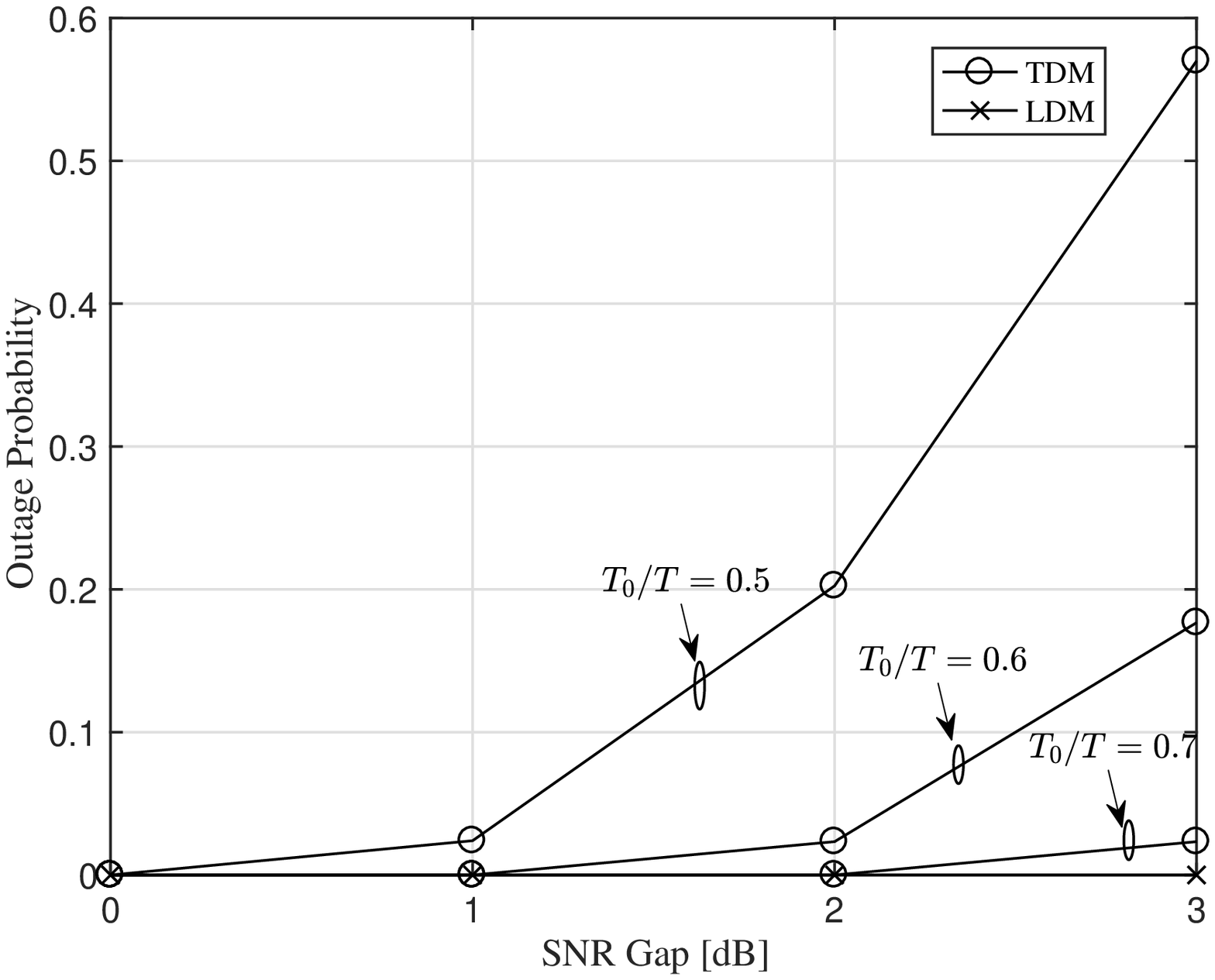}
\label{CDFimperfectCC}
}
\subfigure{
\centering
\includegraphics[scale=0.4]{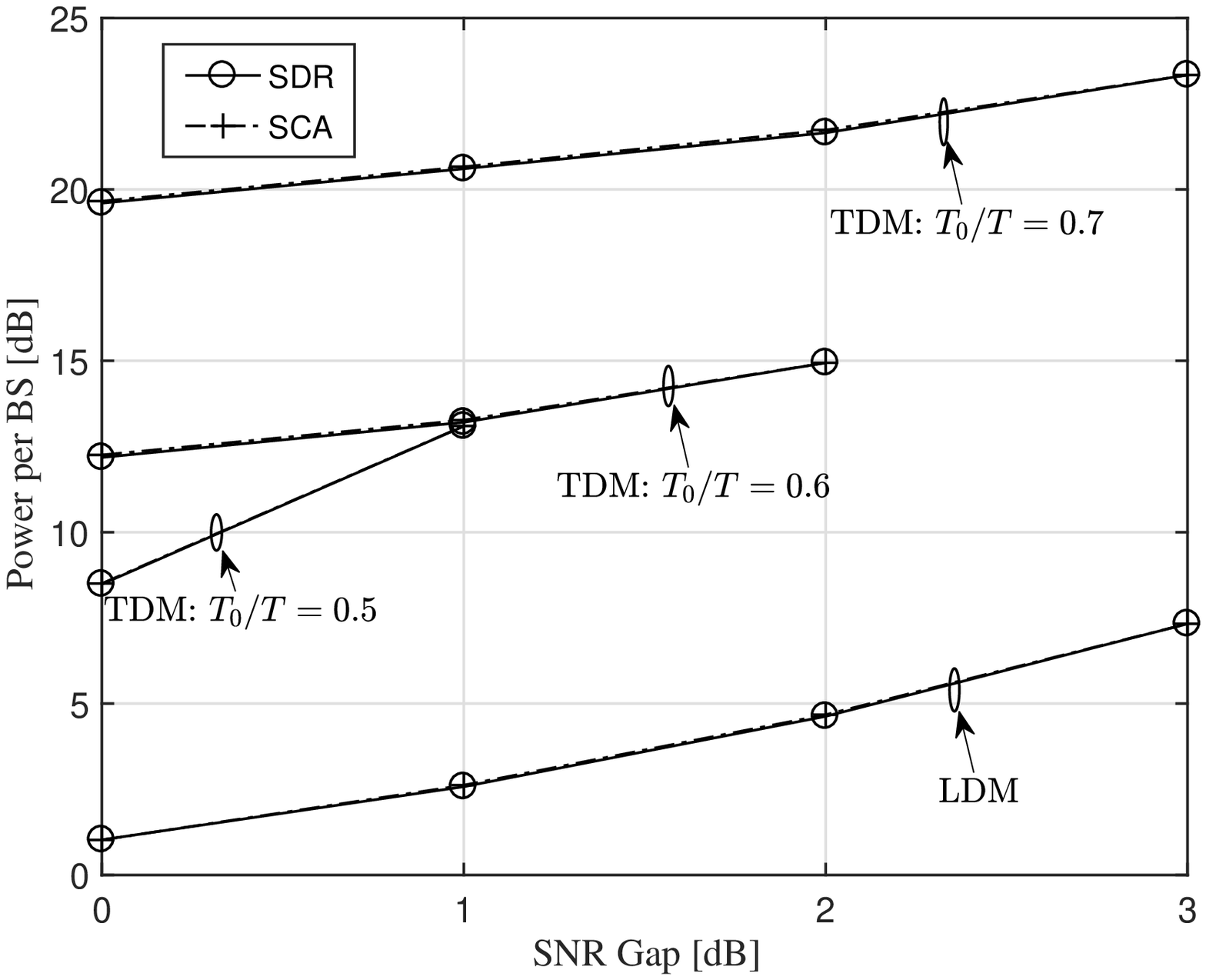}
\label{CDFimperfectCCpower}
}
\caption{Outage probability and power consumption per BS for various values of SNR gap from ideal channel coding with target rates \(R^B\)=3~bps/Hz and \(R^U\)=0.5~bps/Hz.}
\end{figure}

The outage probability versus the SNR gap, measured in dB, is presented in Fig.~\ref{CDFimperfectCC}, while the corresponding transmission power per BS for LDM and TDM are depicted in Fig.~\ref{CDFimperfectCCpower}. It can be observed that the outage probability of TDM significantly increases with the increased SNR gap from perfect channel coding, while the outage probability of LDM remains zero in our setting. In the state-of-the-art terrestrial broadcasting system where $\lambda^{U}=\lambda^{B}=-1~\text{dB}$ are considered as the realistic values for the SNR gaps of the two layers \cite{compare}, although TDM provides acceptable system service availability, the power consumption is found to be about $10~\text{dB}$ higher than LDM, as shown in Fig.~\ref{CDFimperfectCCpower}. It can be further noticed that even when the SNR gap is $3~\text{dB}$ in LDM, the power consumption is still lower than TDM with ideal channel coding.

\subsection{Imperfect CSI}
\begin{figure}[!tp]
\centering
\includegraphics[width=3.5in]{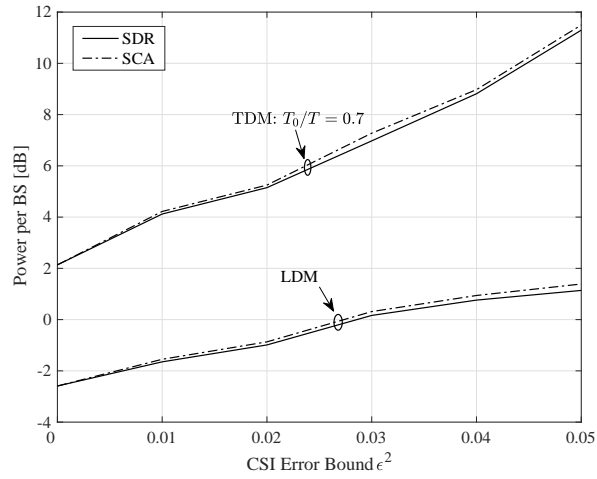}
\caption{Power consumption per BS as a function of CSI error bound $\epsilon^2$ with target rates \(R^B\)=1 bps/Hz and \(R^U\)=1 bps/Hz.}
\label{imperfectcsi}
\end{figure}

We then demonstrate the effect of imperfect CSI on the performance. The channel error covariance matrix is set as $\bm{Q}_{i,n,k} = 1/\epsilon^2 \bm{I}_{M}$, where $\epsilon^2$ is the common CSI error variance for all $\bm{e}_{i,n,k}$'s. It is observed in Fig.~\ref{imperfectcsi} that the power consumption per BS increases for both TDM and LDM systems, with the increase in CSI error variance $\epsilon^2$. It is interesting to note that the minimum required power of TDM increases faster than that of LDM, indicating that TDM is more sensitive to CSI errors compared to LDM. This effect resembles the results encountered with higher unicast rate requirement and more users. In general, LDM outperforms TDM in terms not only of power consumption, but also of robustness against flexible system QoS targets and CSI imperfections.

\begin{figure}[!tp]
\centering
\subfigure{
\centering
\includegraphics[scale=0.4]{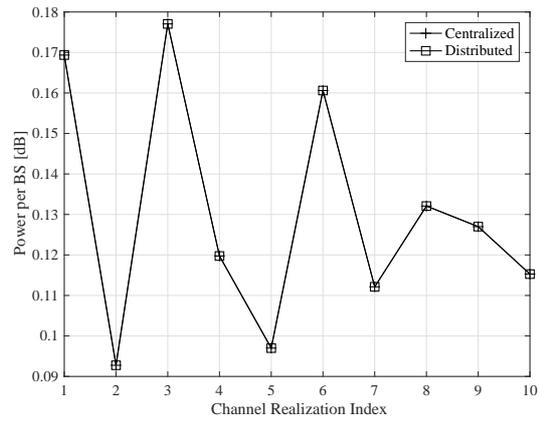}
\label{DD_LDM}
}
\subfigure{
\centering
\includegraphics[scale=0.4]{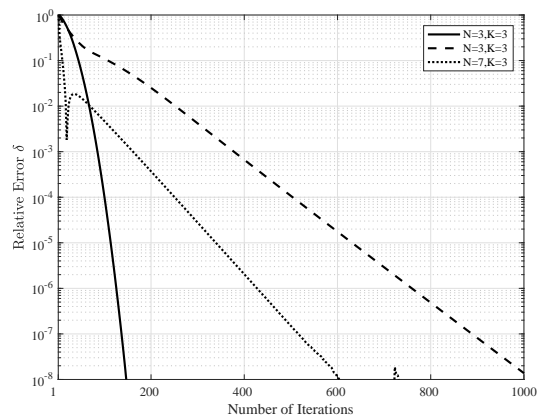}
\label{ADMM}
}
\caption{Convergence of the dual decomposition-based algorithm and relative error within dual ascent iterations for a given SCA subproblem.}
\end{figure}

\subsection{Distributed Implementation}
We first demonstrate that the distributed algorithm can converge to the same optimal solution as the centralized scheme, as shown in Fig.~\ref{DD_LDM}. The centralized solution was obtained by solving the optimization problem in (\ref{eq:opt_LDM_SCA}) by CVX.
The performance of the proposed dual decomposition-based distributed algorithm is studied in Fig.~\ref{ADMM}.
The relative error at the $j$-th iteration of the algorithm in the $\nu$-th SCA loop is computed by $\delta=|p^j-p^*|/p^*$, where $p^j$ denotes the dual ascent solution at iteration $j$, and $p^*$ denotes the optimal solution obtained by CVX in the best precision mode. The appropriate penalty parameters $\rho$ are found empirically to observe fast convergence. Accordingly, Fig.~\ref{ADMM} shows the convergence behaviour of the distributed solution as a function of the number of iterations. It can be seen that for LDM, the algorithm converges fast to achieve an acceptable relative value, say $\delta=10^{-4}$, within 500 iterations for a $N=7$ cell network.

\section{Conclusions}
In this paper, we have analyzed the performance gain of LDM over TDM/FDM as a potential NOMA approach for simultaneous transmission of broadcast and unicast messages over cellular networks. Joint beamforming design and power allocation was formulated as a sum-power minimization problem under distinct QoS constraints for the individual unicast messages and the common broadcast message. The resulting non-convex problem has been tackled by means of SCA and $\rm{S}$-procedure, which provide upper and lower bounds on the optimal solution, respectively.
Our numerical results have shown that the upper and lower bounds are tight, which indicates the near-optimality of the proposed solutions. We have also observed that LDM significantly improves the performance as compared to orthogonal transmission, and that it provides power savings for both the unicast and broadcast transmissions thanks to the larger bandwidth available. We have seen that the benefit of the increased bandwidth available for the broadcast layer outweights the interference caused by unicast transmissions. In the case of imperfect CSI, we have noted that, while increased CSI error adversely affects both LDM and TDM, the increase in minimum required power as a function of the CSI error variance is much faster with TDM compared to LDM, indicating that LDM also provides better robustness against CSI uncertainties commonly experienced in real systems.
A dual decomposition-based distributed solution has also been presented, which facilitates efficient distributed implementation for the LDM technique.

\bibliographystyle{IEEEtran}
\bibliography{IEEEabrv,LDM}

\end{document}